\documentclass[9pt, sigconf, letterpaper]{acmart}

\usepackage{url}
\usepackage[utf8]{inputenc}
\usepackage{xcolor}
\usepackage{amsmath}
\usepackage{xspace}
\usepackage{epsfig}
\usepackage{balance}
\usepackage{booktabs}
\usepackage{tabularx}
\usepackage[font=footnotesize]{subcaption}
\usepackage[font=footnotesize]{caption}
\usepackage{mathtools}
\usepackage{soul}
\usepackage{lipsum}
\usepackage{bm}
\usepackage{enumerate}

\usepackage[acronyms,nonumberlist,nopostdot,nomain,nogroupskip,acronymlists={hidden}]{glossaries}
\newglossary[algh]{hidden}{acrh}{acnh}{Hidden Acronyms}
\glsdisablehyper

\usepackage{tikz}
\usepackage{pgfplots}
\pgfplotsset{compat=newest}
\pgfplotsset{plot coordinates/math parser=false}
\newlength\fheight
\newlength\fwidth
\usetikzlibrary{plotmarks,patterns,decorations.pathreplacing,backgrounds,calc,arrows,arrows.meta,spy,matrix,scopes}
\usepgfplotslibrary{patchplots,groupplots}
\usepackage{tikzscale}

\newacronym{3gpp}{3GPP}{3rd Generation Partnership Project}
\newacronym{4g}{4G}{4th generation}
\newacronym{5g}{5G}{5th generation}
\newacronym{6g}{6G}{6th generation}
\newacronym{5gc}{5GC}{5G Core}
\newacronym{adc}{ADC}{Analog to Digital Converter}
\newacronym{aerpaw}{AERPAW}{Aerial Experimentation and Research Platform for Advanced Wireless}
\newacronym{ai}{AI}{Artificial Intelligence}
\newacronym{aimd}{AIMD}{Additive Increase Multiplicative Decrease}
\newacronym{am}{AM}{Acknowledged Mode}
\newacronym{amc}{AMC}{Adaptive Modulation and Coding}
\newacronym{amf}{AMF}{Access and Mobility Management Function}
\newacronym{aops}{AOPS}{Adaptive Order Prediction Scheduling}
\newacronym{api}{API}{Application Programming Interface}
\newacronym{apn}{APN}{Access Point Name}
\newacronym{aqm}{AQM}{Active Queue Management}
\newacronym{ausf}{AUSF}{Authentication Server Function}
\newacronym{avc}{AVC}{Advanced Video Coding}
\newacronym{awgn}{AGWN}{Additive White Gaussian Noise}
\newacronym{balia}{BALIA}{Balanced Link Adaptation Algorithm}
\newacronym{bbu}{BBU}{Base Band Unit}
\newacronym{bdp}{BDP}{Bandwidth-Delay Product}
\newacronym{ber}{BER}{Bit Error Rate}
\newacronym{bf}{BF}{Beamforming}
\newacronym{bler}{BLER}{Block Error Rate}
\newacronym{brr}{BRR}{Bayesian Ridge Regressor}
\newacronym{bsr}{BSR}{Buffer Status Report}
\newacronym{bs}{BS}{Base Station}
\newacronym{bpsk}{BPSK}{Binary Phase-shift keying}
\newacronym{bss}{BSS}{Business Support System}
\newacronym{ca}{CA}{Carrier Aggregation}
\newacronym{caas}{CaaS}{Connectivity-as-a-Service}
\newacronym{cb}{CB}{Code Block}
\newacronym{cc}{CC}{Congestion Control}
\newacronym{ccid}{CCID}{Congestion Control ID}
\newacronym{cco}{CC}{Carrier Component}
\newacronym{cdd}{CDD}{Cyclic Delay Diversity}
\newacronym{cdf}{CDF}{Cumulative Distribution Function}
\newacronym{cdn}{CDN}{Content Distribution Network}
\newacronym{cir}{CIR}{Channel Impulse Response}
\newacronym{cn}{CN}{Core Network}
\newacronym{codel}{CoDel}{Controlled Delay Management}
\newacronym{comac}{COMAC}{Converged Multi-Access and Core}
\newacronym{cord}{CORD}{Central Office Re-architected as a Datacenter}
\newacronym{cornet}{CORNET}{COgnitive Radio NETwork}
\newacronym{cosmos}{COSMOS}{Cloud Enhanced Open Software Defined Mobile Wireless Testbed for City-Scale Deployment}
\newacronym{cots}{COTS}{Commercial Off-the-Shelf}
\newacronym{cp}{CP}{Control Plane}
\newacronym{cpu}{CPU}{Central Processing Unit}
\newacronym{cqi}{CQI}{Channel Quality Information}
\newacronym{cr}{CR}{Cognitive Radio}
\newacronym{cran}{CRAN}{Cloud \gls{ran}}
\newacronym{crs}{CRS}{Cell Reference Signal}
\newacronym{csi}{CSI}{Channel State Information}
\newacronym{csirs}{CSI-RS}{Channel State Information - Reference Signal}
\newacronym{cu}{CU}{Central Unit}
\newacronym{d2tcp}{D$^2$TCP}{Deadline-aware Data center TCP}
\newacronym{d3}{D$^3$}{Deadline-Driven Delivery}
\newacronym{dac}{DAC}{Digital to Analog Converter}
\newacronym{dag}{DAG}{Directed Acyclic Graph}
\newacronym{darpa}{DARPA}{Defense Advanced Research Projects Agency}
\newacronym{das}{DAS}{Distributed Antenna System}
\newacronym{dash}{DASH}{Dynamic Adaptive Streaming over HTTP}
\newacronym{dc}{DC}{Dual Connectivity}
\newacronym{dccp}{DCCP}{Datagram Congestion Control Protocol}
\newacronym{dce}{DCE}{Direct Code Execution}
\newacronym{dci}{DCI}{Downlink Control Information}
\newacronym{dcl}{DCL}{Dear Colleague Letter}
\newacronym{dctcp}{DCTCP}{Data Center TCP}
\newacronym{dl}{DL}{Downlink}
\newacronym{dmr}{DMR}{Deadline Miss Ratio}
\newacronym{dmrs}{DMRS}{DeModulation Reference Signal}
\newacronym{drlcc}{DRL-CC}{Deep Reinforcement Learning Congestion Control}
\newacronym{drs}{DRS}{Discovery Reference Signal}
\newacronym{du}{DU}{Distributed Unit}
\newacronym{e2e}{E2E}{end-to-end}
\newacronym{ecaas}{ECaaS}{Edge-Cloud-as-a-Service}
\newacronym{ecn}{ECN}{Explicit Congestion Notification}
\newacronym{edf}{EDF}{Earliest Deadline First}
\newacronym{em}{EM}{Electro-Magnetic}
\newacronym{embb}{eMBB}{Enhanced Mobile Broadband}
\newacronym{empower}{EMPOWER}{EMpowering transatlantic PlatfOrms for advanced WirEless Research}
\newacronym{enb}{eNB}{evolved Node Base}
\newacronym{endc}{EN-DC}{E-UTRAN-\gls{nr} \gls{dc}}
\newacronym{epc}{EPC}{Evolved Packet Core}
\newacronym{eps}{EPS}{Evolved Packet System}
\newacronym{es}{ES}{Edge Server}
\newacronym{etsi}{ETSI}{European Telecommunications Standards Institute}
\newacronym[firstplural=Estimated Times of Arrival (ETAs)]{eta}{ETA}{Estimated Time of Arrival}
\newacronym{eutran}{E-UTRAN}{Evolved Universal Terrestrial Access Network}
\newacronym{faas}{FaaS}{Function-as-a-Service}
\newacronym{fapi}{FAPI}{Functional Application Platform Interface}
\newacronym{fcc}{FCC}{Federal Communications Commission}
\newacronym{fdd}{FDD}{Frequency Division Duplexing}
\newacronym{fdm}{FDM}{Frequency Division Multiplexing}
\newacronym{fdma}{FDMA}{Frequency Division Multiple Access}
\newacronym{fed4fire}{FED4FIRE+}{Federation 4 Future Internet Research and Experimentation Plus}
\newacronym{fir}{FIR}{Finite Impulse Response}
\newacronym{fit}{FIT}{Future \acrlong{iot}}
\newacronym{fpga}{FPGA}{Field Programmable Gate Array}
\newacronym{fr2}{FR2}{Frequency Range 2}
\newacronym{fs}{FS}{Fast Switching}
\newacronym{fscc}{FSCC}{Flow Sharing Congestion Control}
\newacronym{ftp}{FTP}{File Transfer Protocol}
\newacronym{fw}{FW}{Flow Window}
\newacronym{ga128}{Ga}{Golay Sequence type A}
\newacronym{ge}{GE}{Gaussian Elimination}
\newacronym{glfsr}{GLFSR}{Galois Linear Feedback Shift Register}
\newacronym{gnb}{gNB}{Next Generation Node Base}
\newacronym{gold}{Gold}{Gold}
\newacronym{gop}{GOP}{Group of Pictures}
\newacronym{gpr}{GPR}{Gaussian Process Regressor}
\newacronym{gpu}{GPU}{Graphics Processing Unit}
\newacronym{gtp}{GTP}{GPRS Tunneling Protocol}
\newacronym{gtpc}{GTP-C}{GPRS Tunnelling Protocol Control Plane}
\newacronym{gtpu}{GTP-U}{GPRS Tunnelling Protocol User Plane}
\newacronym{gtpv2c}{GTPv2-C}{\gls{gtp} v2 - Control}
\newacronym{gw}{GW}{Gateway}
\newacronym{harq}{HARQ}{Hybrid Automatic Repeat reQuest}
\newacronym{hetnet}{HetNet}{Heterogeneous Network}
\newacronym{hh}{HH}{Hard Handover}
\newacronym{hol}{HOL}{Head-of-Line}
\newacronym{hqf}{HQF}{Highest-quality-first}
\newacronym{hss}{HSS}{Home Subscription Server}
\newacronym{http}{HTTP}{HyperText Transfer Protocol}
\newacronym{ia}{IA}{Initial Access}
\newacronym{iab}{IAB}{Integrated Access and Backhaul}
\newacronym{ic}{IC}{Incident Command}
\newacronym{ietf}{IETF}{Internet Engineering Task Force}
\newacronym{ifw}{IFW}{Interference Free Window}
\newacronym{imsi}{IMSI}{International Mobile Subscriber Identity}
\newacronym{imt}{IMT}{International Mobile Telecommunication}
\newacronym{iot}{IoT}{Internet of Things}
\newacronym{ip}{IP}{Internet Protocol}
\newacronym{iq}{IQ}{In-phase and Quadrature}
\newacronym{itu}{ITU}{International Telecommunication Union}
\newacronym{kpi}{KPI}{Key Performance Indicator}
\newacronym{kvm}{KVM}{Kernel-based Virtual Machine}
\newacronym{los}{LOS}{Line-of-Sight}
\newacronym{ls}{LS}{Loosely Synchronised}
\newacronym{lsm}{LSM}{Link-to-System Mapping}
\newacronym{lstm}{LSTM}{Long Short Term Memory}
\newacronym{lte}{LTE}{Long Term Evolution}
\newacronym{lxc}{LXC}{Linux Container}
\newacronym{m2m}{M2M}{Machine to Machine}
\newacronym{mac}{MAC}{Medium Access Control}
\newacronym{manet}{MANET}{Mobile Ad Hoc Network}
\newacronym{mano}{MANO}{Management and Orchestration}
\newacronym{mc}{MC}{Multi-Connectivity}
\newacronym{mcc}{MCC}{Mobile Cloud Computing}
\newacronym{mchem}{MCHEM}{Massive Channel Emulator}
\newacronym{mcs}{MCS}{Modulation and Coding Scheme}
\newacronym{mec}{MEC}{Multi-access Edge Computing}
\newacronym{mec2}{MEC}{Mobile Edge Cloud}
\newacronym{mfc}{MFC}{Mobile Fog Computing}
\newacronym{mi}{MI}{Mutual Information}
\newacronym{mib}{MIB}{Master Information Block}
\newacronym{miesm}{MIESM}{Mutual Information Based Effective SINR}
\newacronym{mimo}{MIMO}{Multiple Input, Multiple Output}
\newacronym{mgen}{MGEN}{Multi-Generator}
\newacronym{ml}{ML}{Machine Learning}
\newacronym{mlr}{MLR}{Maximum-local-rate}
\newacronym[plural=\gls{mme}s,firstplural=Mobility Management Entities (MMEs)]{mme}{MME}{Mobility Management Entity}
\newacronym{mmtc}{mMTC}{Massive Machine-Type Communications}
\newacronym{mmwave}{mmWave}{millimeter wave}
\newacronym{mpdccp}{MP-DCCP}{Multipath Datagram Congestion Control Protocol}
\newacronym{mptcp}{MPTCP}{Multipath TCP}
\newacronym{mr}{MR}{Maximum Rate}
\newacronym{mrdc}{MR-DC}{Multi \gls{rat} \gls{dc}}
\newacronym{mse}{MSE}{Mean Square Error}
\newacronym{mss}{MSS}{Maximum Segment Size}
\newacronym{mt}{MT}{Mobile Termination}
\newacronym{mtd}{MTD}{Machine-Type Device}
\newacronym{mtu}{MTU}{Maximum Transmission Unit}
\newacronym{mumimo}{MU-MIMO}{Multi-user \gls{mimo}}
\newacronym{mvno}{MVNO}{Mobile Virtual Network Operator}
\newacronym{nalu}{NALU}{Network Abstraction Layer Unit}
\newacronym{nas}{NAS}{Network Attached Storage}
\newacronym{nbiot}{NB-IoT}{Narrow Band IoT}
\newacronym{nfv}{NFV}{Network Function Virtualization}
\newacronym{nfvi}{NFVI}{Network Function Virtualization Infrastructure}
\newacronym{nic}{NIC}{Network Interface Card}
\newacronym{nlos}{NLOS}{Non-Line-of-Sight}
\newacronym{now}{NOW}{Non Overlapping Window}
\newacronym{nrdz}{NRDZ}{National Radio Dynamic Zone}
\newacronym{nsf}{NSF}{National Science Foundation}
\newacronym{nsm}{NSM}{Network Service Mesh}
\newacronym[type=hidden]{nr}{NR}{New Radio}
\newacronym{nrf}{NRF}{Network Repository Function}
\newacronym{nsa}{NSA}{Non Stand Alone}
\newacronym{nse}{NSE}{Network Slicing Engine}
\newacronym{nssf}{NSSF}{Network Slice Selection Function}
\newacronym{ntp}{NTP}{Network Time Protocol}
\newacronym{o2i}{O2I}{Outdoor to Indoor}
\newacronym{oai}{OAI}{OpenAirInterface}
\newacronym{oaicn}{OAI-CN}{\gls{oai} \acrlong{cn}}
\newacronym{oairan}{OAI-RAN}{\acrlong{oai} \acrlong{ran}}
\newacronym{oam}{OAM}{Operations, Administration and Maintenance}
\newacronym[plural=\gls{obu}s,firstplural=Onboard Units (OBUs)]{obu}{OBU}{Onboard Unit}
\newacronym{ofdm}{OFDM}{Orthogonal Frequency Division Multiplexing}
\newacronym{olia}{OLIA}{Opportunistic Linked Increase Algorithm}
\newacronym{omec}{OMEC}{Open Mobile Evolved Core}
\newacronym{onap}{ONAP}{Open Network Automation Platform}
\newacronym{onf}{ONF}{Open Networking Foundation}
\newacronym{onos}{ONOS}{Open Networking Operating System}
\newacronym{oom}{OOM}{\gls{onap} Operations Manager}
\newacronym{opnfv}{OPNFV}{Open Platform for \gls{nfv}}
\newacronym[type=hidden]{oran}{O-RAN}{Open \gls{ran}}
\newacronym{orbit}{ORBIT}{Open-Access Research Testbed for Next-Generation Wireless Networks}
\newacronym{os}{OS}{Operating System}
\newacronym{osm}{OSM}{Open Street Map}
\newacronym{oss}{OSS}{Operations Support System}
\newacronym{pa}{PA}{Position-aware}
\newacronym{pase}{PASE}{Prioritization, Arbitration, and Self-adjusting Endpoints}
\newacronym{pawr}{PAWR}{Platforms for Advanced Wireless Research}
\newacronym{pbch}{PBCH}{Physical Broadcast Channel}
\newacronym{pcef}{PCEF}{Policy and Charging Enforcement Function}
\newacronym{pcfich}{PCFICH}{Physical Control Format Indicator Channel}
\newacronym{pcrf}{PCRF}{Policy and Charging Rules Function}
\newacronym{pdcch}{PDCCH}{Physical Downlink Control Channel}
\newacronym{pdcp}{PDCP}{Packet Data Convergence Protocol}
\newacronym{pdsch}{PDSCH}{Physical Downlink Shared Channel}
\newacronym{pdu}{PDU}{Packet Data Unit}
\newacronym{pdp}{PDP}{Power Delay Profile}
\newacronym{pf}{PF}{Proportional Fair}
\newacronym{pgw}{PGW}{Packet Gateway}
\newacronym{phich}{PHICH}{Physical Hybrid ARQ Indicator Channel}
\newacronym{phy}{PHY}{Physical}
\newacronym{pl}{PL}{Path Loss}
\newacronym{pmch}{PMCH}{Physical Multicast Channel}
\newacronym{pmi}{PMI}{Precoding Matrix Indicators}
\newacronym{powder}{POWDER}{Platform for Open Wireless Data-driven Experimental Research}
\newacronym{ppo}{PPO}{Proximal Policy Optimization}
\newacronym{ppp}{PPP}{Poisson Point Process}
\newacronym{prach}{PRACH}{Physical Random Access Channel}
\newacronym{prb}{PRB}{Physical Resource Block}
\newacronym{psnr}{PSNR}{Peak Signal to Noise Ratio}
\newacronym{pss}{PSS}{Primary Synchronization Signal}
\newacronym{pucch}{PUCCH}{Physical Uplink Control Channel}
\newacronym{pusch}{PUSCH}{Physical Uplink Shared Channel}
\newacronym{qam}{QAM}{Quadrature Amplitude Modulation}
\newacronym{qci}{QCI}{\gls{qos} Class Identifier}
\newacronym{qoe}{QoE}{Quality of Experience}
\newacronym{qos}{QoS}{Quality of Service}
\newacronym{qtgui}{QT-GUI}{QT Graphical User Interface}
\newacronym{quic}{QUIC}{Quick UDP Internet Connections}
\newacronym{rach}{RACH}{Random Access Channel}
\newacronym{ran}{RAN}{Radio Access Network}
\newacronym[firstplural=Radio Access Technologies (RATs)]{rat}{RAT}{Radio Access Technology}
\newacronym{rcn}{RCN}{Research Coordination Network}
\newacronym{rec}{REC}{Radio Edge Cloud}
\newacronym{red}{RED}{Random Early Detection}
\newacronym{renew}{RENEW}{Reconfigurable Eco-system for Next-generation End-to-end Wireless}
\newacronym{rf}{RF}{Radio Frequency}
\newacronym{rfc}{RFC}{Request for Comments}
\newacronym{rfr}{RFR}{Random Forest Regressor}
\newacronym{ric}{RIC}{\gls{ran} Intelligent Controller}
\newacronym{rlc}{RLC}{Radio Link Control}
\newacronym{rlf}{RLF}{Radio Link Failure}
\newacronym{rlnc}{RLNC}{Random Linear Network Coding}
\newacronym{rmse}{RMSE}{Root Mean Squared Error}
\newacronym{rnis}{RNIS}{Radio Network Information Service}
\newacronym{rr}{RR}{Round Robin}
\newacronym{rrc}{RRC}{Radio Resource Control}
\newacronym{rrm}{RRM}{Radio Resource Management}
\newacronym{rru}{RRU}{Remote Radio Unit}
\newacronym{rs}{RS}{Remote Server}
\newacronym{rsrp}{RSRP}{Reference Signal Received Power}
\newacronym{rsrq}{RSRQ}{Reference Signal Received Quality}
\newacronym{rss}{RSS}{Received Signal Strength}
\newacronym{rssi}{RSSI}{Received Signal Strength Indicator}
\newacronym{rsu}{RSU}{Road-Side Unit}
\newacronym{rtt}{RTT}{Round Trip Time}
\newacronym{ru}{RU}{Radio Unit}
\newacronym{rw}{RW}{Receive Window}
\newacronym{rx}{RX}{Receiver}
\newacronym{s1ap}{S1AP}{S1 Application Protocol}
\newacronym{sa}{SA}{standalone}
\newacronym{sack}{SACK}{Selective Acknowledgment}
\newacronym{sap}{SAP}{Service Access Point}
\newacronym{sc2}{SC2}{Spectrum Collaboration Challenge}
\newacronym{scef}{SCEF}{Service Capability Exposure Function}
\newacronym{sch}{SCH}{Secondary Cell Handover}
\newacronym{scoot}{SCOOT}{Split Cycle Offset Optimization Technique}
\newacronym{sctp}{SCTP}{Stream Control Transmission Protocol}
\newacronym{sdap}{SDAP}{Service Data Adaptation Protocol}
\newacronym{sd}{SD}{Standard Deviation}
\newacronym{sdk}{SDK}{Software Development Kit}
\newacronym{sdm}{SDM}{Space Division Multiplexing}
\newacronym{sdma}{SDMA}{Spatial Division Multiple Access}
\newacronym{sdn}{SDN}{Software-defined Networking}
\newacronym{sdr}{SDR}{Software-defined Radio}
\newacronym{seba}{SEBA}{SDN-Enabled Broadband Access}
\newacronym{sgsn}{SGSN}{Serving GPRS Support Node}
\newacronym{sgw}{SGW}{Service Gateway}
\newacronym{si}{SI}{Study Item}
\newacronym{sib}{SIB}{Secondary Information Block}
\newacronym{sinr}{SINR}{Signal to Interference plus Noise Ratio}
\newacronym{sip}{SIP}{Session Initiation Protocol}
\newacronym{siso}{SISO}{Single Input, Single Output}
\newacronym{sla}{SLA}{Service Level Agreement}
\newacronym{sm}{SM}{Saturation Mode}
\newacronym{smf}{SMF}{Session Management Function}
\newacronym{smo}{SMO}{Service Management and Orchestration}
\newacronym{sms}{SMS}{Short Message Service}
\newacronym{smsgmsc}{SMS-GMSC}{\gls{sms}-Gateway}
\newacronym{snr}{SNR}{Signal-to-Noise-Ratio}
\newacronym{son}{SON}{Self-Organizing Network}
\newacronym{sptcp}{SPTCP}{Single Path TCP}
\newacronym{srb}{SRB}{Service Radio Bearer}
\newacronym{srn}{SRN}{Standard Radio Node}
\newacronym{srs}{SRS}{Sounding Reference Signal}
\newacronym{ss}{SS}{Synchronization Signal}
\newacronym{sss}{SSS}{Secondary Synchronization Signal}
\newacronym{st}{ST}{Spanning Tree}
\newacronym{svc}{SVC}{Scalable Video Coding}
\newacronym{tb}{TB}{Transport Block}
\newacronym{tcp}{TCP}{Transmission Control Protocol}
\newacronym{tdd}{TDD}{Time Division Duplexing}
\newacronym{tdm}{TDM}{Time Division Multiplexing}
\newacronym{tdma}{TDMA}{Time Division Multiple Access}
\newacronym{tfl}{TfL}{Transport for London}
\newacronym{tfrc}{TFRC}{TCP-Friendly Rate Control}
\newacronym{tft}{TFT}{Traffic Flow Template}
\newacronym{tgen}{TGEN}{Traffic Generator}
\newacronym{tip}{TIP}{Telecom Infra Project}
\newacronym{tm}{TM}{Transparent Mode}
\newacronym{to}{TO}{Telco Operator}
\newacronym{toa}{ToA}{Time of Arrival}
\newacronym{tr}{TR}{Technical Report}
\newacronym{trp}{TRP}{Transmitter Receiver Pair}
\newacronym{ts}{TS}{Technical Specification}
\newacronym{tti}{TTI}{Transmission Time Interval}
\newacronym{ttt}{TTT}{Time-to-Trigger}
\newacronym{tx}{TX}{Transmitter}
\newacronym{uas}{UAS}{Unmanned Aerial System}
\newacronym{uav}{UAV}{Unmanned Aerial Vehicle}
\newacronym{udm}{UDM}{Unified Data Management}
\newacronym{udp}{UDP}{User Datagram Protocol}
\newacronym{udr}{UDR}{Unified Data Repository}
\newacronym{ue}{UE}{User Equipment}
\newacronym{uhd}{UHD}{\gls{usrp} Hardware Driver}
\newacronym{ul}{UL}{Uplink}
\newacronym{um}{UM}{Unacknowledged Mode}
\newacronym{uml}{UML}{Unified Modeling Language}
\newacronym{upa}{UPA}{Uniform Planar Array}
\newacronym{upf}{UPF}{User Plane Function}
\newacronym{urllc}{URLLC}{Ultra Reliable and Low Latency Communication}
\newacronym{usa}{U.S.}{United States}
\newacronym{usim}{USIM}{Universal Subscriber Identity Module}
\newacronym{usrp}{USRP}{Universal Software Radio Peripheral}
\newacronym{utc}{UTC}{Urban Traffic Control}
\newacronym{vim}{VIM}{Virtualization Infrastructure Manager}
\newacronym{vm}{VM}{Virtual Machine}
\newacronym{vnf}{VNF}{Virtual Network Function}
\newacronym{volte}{VoLTE}{Voice over \gls{lte}}
\newacronym{voltha}{VOLTHA}{Virtual OLT HArdware Abstraction}
\newacronym{vr}{VR}{Virtual Reality}
\newacronym{vran}{vRAN}{Virtualized \gls{ran}}
\newacronym{vss}{VSS}{Video Streaming Server}
\newacronym{wbf}{WBF}{Wired Bias Function}
\newacronym{wf}{WF}{Wired-first}
\newacronym{wi}{WI}{Wireless InSite}
\newacronym{wlan}{WLAN}{Wireless Local Area Network}
\newacronym{pnf}{PNF}{Physical Network Function}
\newacronym{drl}{DRL}{Deep Reinforcement Learning}
\newacronym{mtc}{MTC}{Machine-type Communications}
\newacronym{v2x}{V2X}{Vehicle-to-everything}
\newacronym{cast}{\textit{CaST}}{Channel emulation generator and Sounder Toolchain}

\usepackage{algorithm}
\usepackage{algorithmicx}
\usepackage{algcompatible}
\usepackage[noend]{algpseudocode}
\usepackage{booktabs}
\usepackage{soul}

\title[CaST: A Toolchain for Creating and Characterizing Realistic Wireless Network Emulation Scenarios]{CaST: A Toolchain for Creating and Characterizing Realistic Wireless Network Emulation Scenarios}

\author[D. Villa, M. Tehrani-Moayyed, P. Johari, S. Basagni, T. Melodia]{Davide Villa, Miead Tehrani-Moayyed, Pedram Johari, Stefano Basagni, Tommaso Melodia}
\affiliation{%
  \institution{Institute for the Wireless Internet of Things, Northeastern University, Boston MA, 02115, USA}
  \city{}
  \state{}
  \country{}
  }
\email{{villa.d, tehranimoayyed.m, p.johari, s.basagni, melodia}@northeastern.edu}

\thanks{This work was partially supported by the U.S.\ National Science Foundation under grant CNS-1925601, and by the U.S.\ Department of Transportation, Federal Highway Administration} 

\begin{abstract}
Large-scale wireless testbeds are being increasingly used in developing and evaluating new solutions for next generation wireless networks. 
Among others, high-fidelity FPGA-based emulation platforms have unique capabilities for faithfully modeling real-world wireless environments in real-time and at scale, while guaranteeing repeatability. 
However, the reliability of the solutions tested on emulation platforms heavily depends on the precision of the emulation process, which is often overlooked. 
To address this unmet need in wireless network emulator-based experiments, in this paper we present \textit{CaST}, a \emph{Channel emulation generator and Sounder Toolchain} for creating and characterizing realistic wireless network scenarios with high accuracy.
\textit{CaST} consists of (i) a framework for creating mobile wireless scenarios from ray-tracing models for FPGA-based emulation platforms, and (ii) a containerized Software Defined Radio-based channel sounder to precisely characterize the emulated channels. 
We demonstrate the use of \textit{CaST} by designing, deploying and validating multi-path mobile scenarios on Colosseum, the world's largest wireless network emulator. 
Results show that \textit{CaST} achieves $\leq 20\:\mathrm{ns}$ accuracy in sounding Channel Impulse Response tap delays, and $0.5\:\mathrm{dB}$ accuracy in measuring tap gains.
\end{abstract}



\begin{document}


\maketitle


\glsresetall
\glsunset{usrp}
\glsunset{uhd}

\section{Introduction}
\label{sec:intro}

The wireless networking industry is experiencing a tremendous growth, as shown by the standardization of~5th generation (5G) technologies and by the vigorous rise of~6G~\cite{giordani2020toward}. 
The need for faster, more reliable, and low-latency wireless technologies is providing a major motivation for researchers to define and develop hosts of new solutions for next generation wireless networks.
In parallel, there has been significant interests and promising advancements in the use of \gls{ai} and data-driven methods to address complex problems in the wireless telecommunications domain that are envisioned to largely replace the traditional model-driven techniques in the years to come.

Needless to say, developing new AI-driven telecommunication solutions requires extensive testing in a variety of environments to demonstrate desired performance. 
However, it is costly and often unfeasible to develop and debug new solutions on large and diverse real-world experimental setups. 
In this context, large-scale wireless emulation platforms have been widely demonstrated to be a valuable resource to design, develop, and validate new applications in quasi-realistic environments, at scale, and with a variety of different topologies, traffic scenarios, and channel conditions~\cite{bonati2021colosseum,sichitiu2020aerpaw,dandekar2019grid}. These network emulators can represent virtually any real-world scenario, also enabling repeatability of experiments.

The reliability of the solutions developed in emulated platforms depends greatly on the precision of the emulation process and of the models of the environment. 
Most channel emulators are based on \gls{fir} filters with pre-defined complex-valued taps that represent the characteristics of the channel, as the \gls{cir} in the baseband. 
Additional complexity is added by multi-path scenarios with mobile nodes, such as \gls{v2x} and \gls{uav} communications, which are relevant to next generation networking.
Modeling these scenarios is no easy feat.
Trade-offs and limitations imposed by the design of the channel emulator, and impairments from hardware-in-the-loop features may compromise the accuracy of the channel modeling process and consequently of the emulated RF environment.
We observe, however, that the validation of the emulated channel characterization is often neglected and considered to be true as defined by the model parameters. 
It is therefore necessary to appropriately evaluate the implementation of the channel models, measure potential emulation errors, and to use the finding to further develop corrective measures to compensate for deviations from desired and expected behaviors.

%
Validating emulation and simulation models of wireless scenarios has been done before. 
Patnaik et al., for instance, investigate the difference between an \gls{fir} filter response and its simulated twin~\cite{patnaik2014implementation}. 
Ju and Rappaport consider spatial consistency to better simulate channel impairments 
in a mmWave channel simulator~\cite{ju2018simulating}. 
Researchers have also exploited ray-tracing to include mobility in emulated channels~\cite{bilibashi2020dynamic, oliveria2019ray, patnaik2014implementation}. 
To the best of our knowledge, however, these works concern very specific scenarios and experiments. 

In this paper, we develop a general, open-source and fully customizable \gls{sdr}-based toolchain to streamline the generation and validation of virtually any type of wireless environment that can be implemented into wireless network emulators at scale.
Our \emph{\gls{cast}} brings to the wireless network emulator landscape a fully open, virtualized and software-based channel generator and sounder toolchain. 
Specifically, our contributions are as follows:

\begin{enumerate}[i.]

    \item We design and develop a streamlined framework to create realistic wireless scenarios with mobility support and based on precise ray-tracing methods for \gls{fir}-based emulation platforms such as Colosseum~\cite{bonati2021colosseum}. 
    
    \item We develop an \gls{sdr}-based channel sounder to precisely characterize emulated RF channels. The sounder framework is fully containerized, scalable, and automated to capture gains and delays of the channel \gls{cir} taps.
    
    \item We test and validate multi-path mobile scenarios on Colosseum, showing that \gls{cast} achieves up to $20\:\mathrm{ns}$ accuracy in sounding \gls{cir} tap delays, and $0.5\:\mathrm{dB}$ accuracy in measuring tap gains.

\end{enumerate}

\noindent
In addition, \gls{cast} is openly available to the whole research community.\footnote{\label{foot:castgithub}\url{https://github.com/wineslab/cast}} \gls{cast} enables wireless research community to design, develop, and test their own new realistic channel models resembling accurate wireless environment of their choice by utilizing a variety of input sources, e.g., ray-tracing software, statistical channel modeling tools, and real-world measurements, among others.

The rest of the paper is organized as follows. 
Sections~\ref{sec:sounding} and~\ref{sec:mobility} describe the two main components of \gls{cast}, sounding and channel mobility generation, respectively, with an overview of the emulation process of Colosseum. 
Section~\ref{sec:results} provides the validation and experimental results of \gls{cast}.
Section~\ref{sec:conclusions} concludes our work.

\section{Channel Sounding}
\label{sec:sounding}

Wireless network emulators such as Colosseum~\cite{bonati2021colosseum} do not involve communication over-the-air.
The RF medium is emulated by means of \gls{fir}-based filter taps in the baseband. 
%
In this context, the main purpose of channel sounding is validating channel emulation traces rather than acquiring physical environment characteristics. 
Consequently, the results of the sounding are compared with the original desired channel model that is given to the network emulator as input. 
These input channel models can be created by different tools, including statistical channel modeling software, ray-tracers, or real-world measurements.

\subsection{The Colosseum emulation system}
\label{sec:colosseum}

Channel emulators are efficient tools for repeatable experiments of solutions for wireless networked systems. 
Remotely and publicly accessible emulators include AERPAW, for UAV-enabled applications~\cite{sichitiu2020aerpaw}, Drexel Grid, an indoor testbed coupling physical \gls{sdr} and virtual nodes~\cite{dandekar2019grid}, and Colosseum, the world largest wireless network emulator~\cite{bonati2021colosseum}, which we use to develop and test \gls{cast}.
Colosseum is a testbed environment with hardware-in-the-loop used to deterministically model real-world RF scenarios.
It provides an open and programmable experimental platform for wireless telecommunication research at scale. 
The Colosseum \gls{mchem} is built around a network of high-performance \gls{fpga} on the ATCA-3671 boards and Ettus Research \gls{usrp} X310 radios. It supports 256 fully connected bidirectional RF channels with $80\:\mathrm{MHz}$ of instantaneous bandwidth with an RF range between $10\:\mathrm{MHz}$ to $6\:\mathrm{GHz}$. The array of Ettus Research \gls{usrp} X310 High Performance \gls{sdr}, each equipped with two UBX-160-LP RF daughter cards, is used as the RF front-end, while ATCA-3671 devices are used to perform the channel computations. The \gls{usrp} X310s are programmed using GNU Radio, \gls{uhd} and Verilog code, while the ACTA-3671 is programmed using BEEcube Platform Studio along with custom Verilog code. With 128 nodes (256 radio front-ends), Colosseum can emulate up to 65K fully independent channels between each transmitter-receiver pair.
Colosseum provides a unique and ideal environment for researchers to create, emulate, run, and test a large amounts of different experiments and real use-case scenarios, allowing a fertile ground where to develop and validate \gls{cast}.

%
%
%

\subsection{\gls{cast} channel sounding architecture}

\begin{figure}[htbp]
    \centering
    \includegraphics[width=.95\columnwidth]{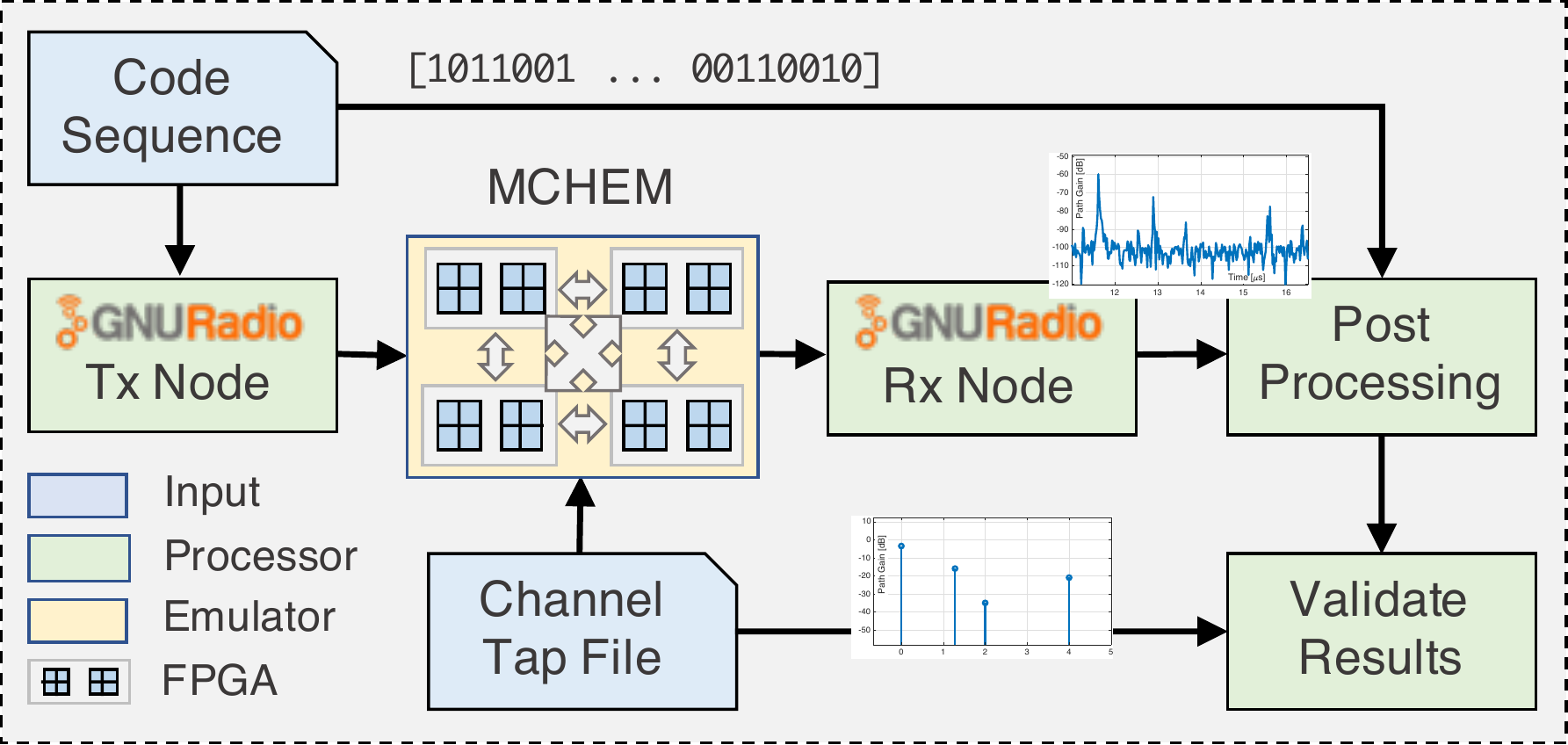}
    \caption{\gls{cast} channel sounding workflow block diagram.}
    \label{fig:soundingbd}
\end{figure}

The \gls{cast} channel sounder consists of two types of nodes: a transmitter, which sends a known sequence, and one or more receivers.
The nodes are developed using the GNU Radio open source \gls{sdr} development toolkit, which enables the design and implementation of software radios through signal processing blocks via hardware or software in-the-loop. 
The transmitted sequence and the received data are stored and then post-processed using MATLAB and Python to obtain channel characterization parameters. Finally, the results are analyzed and compared with the original \gls{cir}.
The corresponding workflow block diagram is shown in Figure~\ref{fig:soundingbd}.

\textbf{Transmitter node.} The transmitter node is designed as a GNU Radio flowgraph with the main duty to build the sequence code and to send it to the receiver via \gls{mchem}. The sending signal is built upon a \textit{Vector Source} block that streams a vector based on an input sequence code and repeats it indefinitely. We consider a \gls{bpsk} modulation where the real part is given by the vector source and transformed in a series of $\{-1,1\}$, while the imaginary part is given by a null source. This stream of data can be displayed through a series of \gls{qtgui} blocks to show time, frequency, and constellation plots of the signal, and is input to a \textit{UHD: USRP Sink} block. The latter block connects the GNU Radio software to a \gls{usrp} hardware device and streams the data with different parameters, e.g., clock source, sample rate, and center frequency. The signal is then transmitted by the \gls{usrp} to \gls{mchem}.

\textbf{Receiver node.}
The receiver node receives and stores the data from \gls{mchem}. The signal is received via a \textit{UHD: USRP Source} block that streams samples from a \gls{usrp} hardware device to the GNU Radio software. The \gls{usrp} acts as a receiver with appropriate parameters, e.g., clock source, sample rate, and center frequency. The received data can be displayed through a series of \gls{qtgui} blocks, and are saved using a \textit{File Sink} block for further post-processing.
It is worth noting that since we operate in a controlled and close environment without any over-the-air transmissions, there is no need to perform any additional control or filtering of the output bandwidth of the transmitted signal to comply with \gls{fcc} regulations. Moreover, \gls{mchem} filters the unwanted signals that are received, leaving only the frequencies of a particular scenario that is running. Thanks to these unique aspects of channel emulators, the transmitter and receiver nodes can be composed of simple yet efficient structures that optimize the channel sounding process and eliminate undesired artifacts.

\subsection{Data processing}
The received data are processed to obtain two main performance metrics: \gls{cir} or $h(t)$, and \gls{pl} or $p(t)$. The \gls{cir} is useful to understand how well the channel reacts in correlation with a given input and reflects the \gls{toa} of the multipath components of a transmitted signal. The \gls{pl} gives information on the intensity of the received signal power on each multipath as a function of time delay, as well as its attenuation. These are fundamental components that need to be considered in the design of a telecommunication system and useful metrics to validate the accuracy of the channel emulation.

Let $c(t)$ be the known code sequence of $N\:\mathrm{bits}$ used by the transmitter node, and $s^{IQ}(t)$ the modulated transmitted sequence with its \gls{iq} components. Similarly, let $r^{IQ}(t)$ be the raw \gls{iq} components stored by the receiver node.
The \gls{cir} \gls{iq} components can be computed by separately correlating $r^{I}(t)$ and $r^{Q}(t)$ of the received data with the $I$ or $Q$ components of $s(t)$ divided by the inner product of the modulated transmitted sequence with its transpose, as shown in Equations~\ref{eq:hicorr} and \ref{eq:hqcorr}, respectively~\cite{garcia2010comparison}:
\begin{equation}
    \label{eq:hicorr}
    h^{I}(t) = (r^{I}(t) \otimes s^{I}(t))/(s^{I^T}(t) * s^{I}(t))
\end{equation}
\begin{equation}
    \label{eq:hqcorr}
    h^{Q}(t) = (r^{Q}(t) \otimes s^{Q}(t))/(s^{Q^T}(t) * s^{Q}(t))
\end{equation}
where $\otimes$ be the cross-correlation between two discrete-time sequences $x$ and $y$~\cite{buck2002computer} which measures the similarity between $x$ and shifted (lagged) repeated copies of $y$ as a function of the lag.
Note that if the considered modulation is \gls{bpsk}, the denominator will be equal to the length of  $c(t)$.
The amplitude of the \gls{cir} can be obtained by Equation~\ref{eq:habs}:
\begin{equation}
    \label{eq:habs}
    |h(t)| = \sqrt{(h^{I}(t))^2 + (h^{Q}(t))^2}
\end{equation}
and the magnitude of the path gains can be calculated as Equation~\ref{eq:pgdb}:
\begin{equation}
    \label{eq:pgdb}
    G_p(t) [dB] = 20log_{10}(|h(t)|) - P_t - G_t - G_r
\end{equation}
where $P_t$ is the power of the transmitted sequence, and $G_t$ and $G_r$ are the gain of transmitter and receiver amplifiers all in $dB$, respectively.  
The \gls{pl} of each multipath can be retrieved by looking at the maximums of $G_p(t)$ in a certain window where the signal of that multipath is received.

\section{Mobility in RF Scenarios}
\label{sec:mobility}

The Colosseum \gls{mchem} supports time-variant channels with minimum coherence time of $1\:\mathrm{ms}$. These channels are read from a tap file that consists of \gls{fir} complex coefficient values for each pair of nodes in a scenario captured at every $1\:\mathrm{ms}$ for the entire duration of any particular scenario. To introduce mobility in Colosseum, we generate a tap file that contains the time-variant \gls{cir}. To facilitate supporting various platform inputs to Colosseum, we divide the mobile scenario generation process into two tasks: (i) scenario generator toolchain; and (ii) mobile channel simulator. The toolchain installs the scenario on Colosseum, incorporates the RF channels data and the traffic metadata into the scenario, and assigns a unique scenario ID. On the other hand, the mobile channel simulator estimates the channels between the mobile nodes using \gls{em} ray-tracing simulator, and implements the movement of the radio nodes in the ray-tracing RF environment.
The mobility simulator is implemented on top of a commercial ray-tracing simulator, namely, \gls{wi}~\cite{WI}, consisting of two steps: (i) sampling the mobile channel using the ray-tracer, and (ii) parsing the ray-tracing outputs to extract the channels for each time instant of emulation. These steps are followed by a channel approximation process that is required to adapt the output channels for emulation~\cite{tehrani2021creating}.



\subsection{V2X scenario in Tampa, FL: A use-case}
\label{sec:usecase_scenario}

\begin{figure}[hbp]
\centering
    \includegraphics[width=0.95\linewidth]{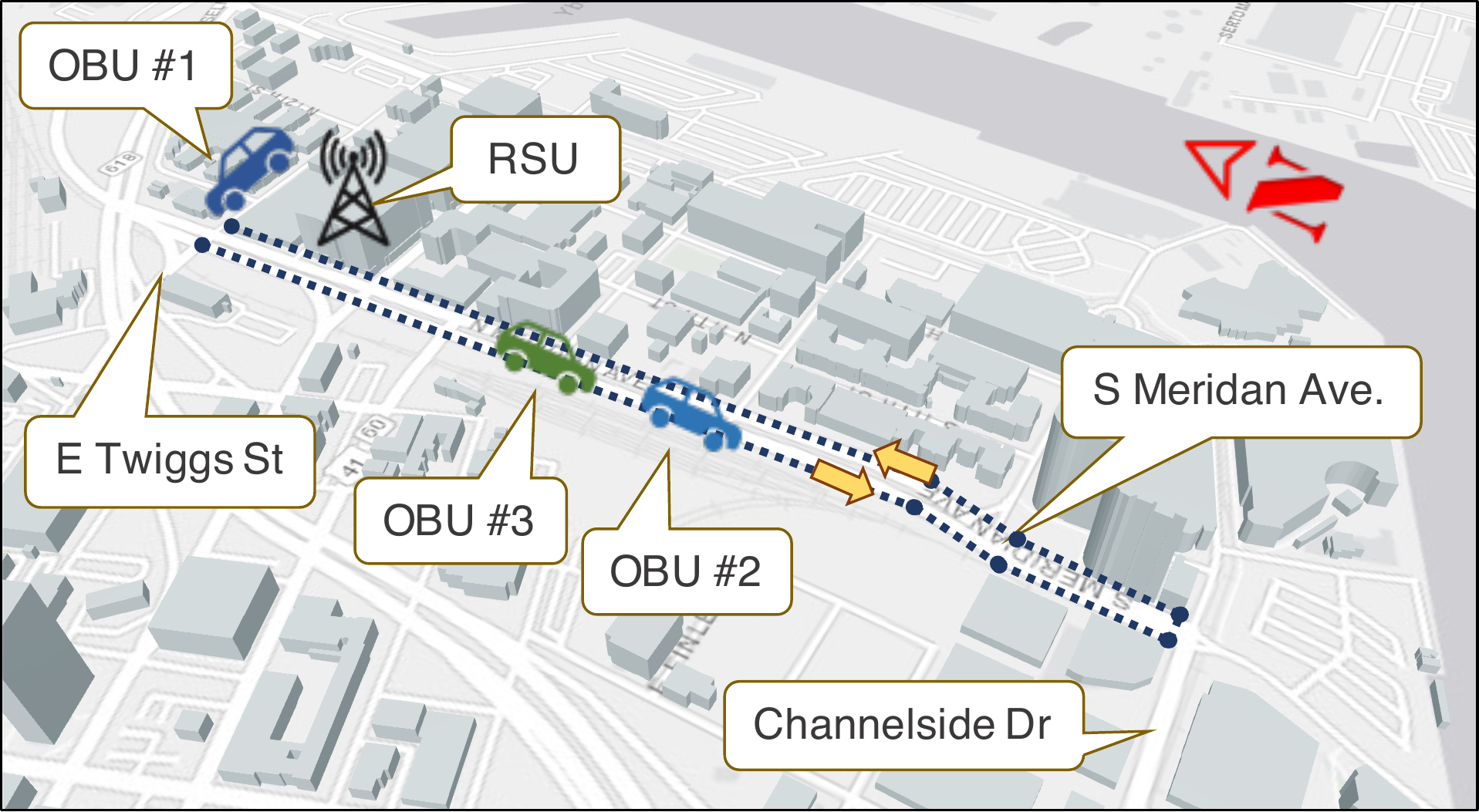}
    \caption{Tampa, FL, V2X scenario simulation environment in WI.}
    \label{fig:Tampa_scenario}
\end{figure}

We consider a scenario around the Tampa Hillsborough Expressway in Tampa, FL.
In order to simulate the scenario in \gls{wi}, we need a 3D model of the wireless environment. 
We obtained such a model from \gls{osm} in XML format and converted it to an STL file, which is supported by \gls{wi}.
The resulting scenario is depicted in Figure~\ref{fig:Tampa_scenario}.

In this scenario, we consider a \gls{v2x} model with four nodes.
One is a \gls{rsu} mounted on the traffic light at the intersection of E.\ Twiggs St.\ and N.\ Meridan Ave.
The other three nodes are \glspl{obu} installed on three vehicles: one is stationary and parked in the parking lot of the Tampa Expressway Authority; the other two vehicles are following each other at a constant speed of $25\:\mathrm{Mph}$ on Meridan Avenue, from E.\ Twiggs St.\ to Channelside Dr., and then back to E.\ Twiggs St. (Figure~\ref{fig:Tampa_scenario}). 
The radio parameters of the nodes are listed in Table~\ref{table:Tampa_wirelessParam}.

\begin{table}[hbp]
\centering
\caption{Wireless parameters for the Tampa simulation scenario.}
\begin{tabular}{@{}ll@{}}
\toprule
\textbf{Parameters} & \textbf{Values for \gls{v2x}} \\
\midrule
Carrier frequency & 5.915 [GHz] \\
Signal bandwidth & 20 [MHz] \\
Transmit power & 20 [dBm] \\
Antenna pattern & Omnidirectional \\
Antenna gain & 5 [dBi] \\
Antenna Height & \gls{rsu}: 16 [ft], \glspl{obu}: 5 [ft] \\
Ambient noise density & -172.8 [dBm/Hz] \\ \bottomrule
\end{tabular}
\label{table:Tampa_wirelessParam}
\end{table}



\subsection{Sampling mobile channels}

{\textbf{Sampling interval and spatial spacing.}}~The effect of mobility is captured by sampling the channels with a predefined sampling time interval of $T_s$ while the transmitter or receivers are moving. This channel sampling concept can be implied in the ray-tracing simulation by spatially sampling the trajectory of the mobile nodes in the scenario with a spacing of $D_i=V_i.T_s$, where $V_i$ is the velocity of the $i^{th}$ mobile node. 

For mobility simulation, it is important to consider spatial consistency, which means that the mobile nodes will experience a similar scattering environment with smooth channel transitions due to the motion. We note that spatial consistency does not deal with the small-scale correlation of received power levels, but rather focuses on providing a consistent and correlated scattering environment that a mobile node experiences~\cite{ju2018simulating}. We consider the coherence distance of $15\:\mathrm{m}$ recommended in \gls{3gpp} Release 14~\cite{3gppModel}, which inherently assures the spatial consistency since we use ray-tracing of the physical environment to simulate the spatial and time evolution, so that the generated parameters of two close locations are highly correlated.

{\textbf{Defining trajectory.}}~The spatial sampling concept can be implemented in \gls{wi} using the ``route'' or  ``trajectory'' type for transmitters and receivers, where the user can define the trajectory of the mobile nodes in the \gls{wi} GUI. This trajectory is defined by determining some control points to specify the start, the end, and where the movement direction is changed. Then, \gls{wi} fills the trajectory with multiple transceiver points and the user can set the spacing between these points. To mimic the channel sampling in real world, we set the spacing with the value calculated as $D = V.T_s$, where $Ts$ is constant for all the mobile nodes, hence the spatial sampling of each node is determined by its velocity.
In our use-case scenario, we consider $V=25\:\mathrm{Mph}$ for mobile vehicles and we set the sampling interval time to $447\:\mathrm{ms}$, hence $D=5\:\mathrm{m}$ which guarantees the spatial consistency. Given the total length of the path $\sim2\:\mathrm{Km}$, \gls{wi} generates $N_{s}=391$ samples along the trajectory shown with dark blue dots/line in Figure~\ref{fig:Tampa_scenario}.

\subsection{Parsing ray-tracing outputs}

In the next step, we parse the spatial samples and the valid channels per time from the ray-tracing outputs to extract synchronized channels between the samples of mobile  and other stationary nodes. \gls{wi} runs the ray-tracing process for a total number of channels between each pair of transceivers $N_{ch}=(N_{nodes})^2$, and will store the results at each timestamp, i.e., a total of $N_{s}.N_{ch}$ channel realizations.

\gls{wi} stores the ray-tracing output of each individual transmitter in a separate file which represents the channels between that individual transmitter and all the receivers in the scenario. 
Algorithm~\ref{alg:mobility_simulator} shows the parsing of the channel outputs between each pair of nodes at all timestamps for the entire duration of a scenario~$T_{total}$. 



\begin{algorithm}[hbp]
\small
\caption{Mobility simulator algorithm.}
\label{alg:mobility_simulator}
\begin{algorithmic}[1]
{
\State{\textit{Number of samples, $N_{s} = \frac{T_{total}-1}{T_s}+1$}}
\State{\textit{$channel$= struct (gain, ToA, AoAs, AoDs)}}
\State{\textit{Channel matrix \textbf{CH}: 3D matrix ($N_{node}$, $N_{node}$, $N_s$) of channel}}
\State{\textit{$time$= time evolution of simulation}}

\For{ sample $s = 1$ to $N_s$}

    \For{ each TX, $i$ }
    \For{ each RX, $j$ }
    
    \If{$V_i = 0 \And s > 1$}
    \State{\textbf{CH}(i,j,s)=\textbf{CH}(i,j,1)}
    \State{\textbf{continue}}
    \EndIf
    
    \State{find TX($i$) sample, $x$ = min($s$, max TX($i$))}
    \State{$\triangleright$ Read ray-tracing output file TX($i$) for $x$}
    \State{find RX($j$) sample, $y$ = min($s$, max RX($j$))}
    \State{$\triangleright$ channel = Extract channel RX($j$) for $y$}
    \State{\textbf{CH} (i,j,s) = channel}
    \State{$time(s) = (s-1) \times T_s$}
    \EndFor
    \EndFor
    
\EndFor
\state{\textbf{Output:} $time$, \textbf{CH}}
}
\end{algorithmic}
\end{algorithm}

The output of the \gls{cast} mobility simulator is a 3D matrix structure that consists of $N_s$ 2D matrix pages. Each page stores the channels between the transmitters in rows and receivers in columns.

\subsection{Time-variant multi-path parameters}

We consider the temporal characteristic of the wireless channel, as an \gls{fir} filter, where the \gls{cir} varies in time and can be expressed as Equation~\ref{eq:time_variant_cir}. $N_t$ is the number of paths at time $t$, $c_i$ is the $i^{th}$ path gain coefficient, and $\tau_i$ is the \gls{toa} of the $i^{th}$ path, which both vary in time. Further, the path gain coefficient is a complex number which carries the magnitude, $a_i$ and phase shift $\varphi_i$ of the $i^{th}$ path (Equation~\ref{eq:cir_coefficient}).
\begin{equation}
    \label{eq:time_variant_cir}
    h(t,\tau) =
    \sum_{i = 1}^{N_t} \Tilde{c_i}(t) . \delta(t-\tau_i(t))
\end{equation}

\begin{equation}
    \label{eq:cir_coefficient}
    \Tilde{c_i}(t) = a_i(t) . e^{j\varphi_{i}(t)}
\end{equation}

We obtain the time-variant \gls{cir} from the estimated parameters of channel paths for each of the valid channel sampled at time instants $t$ from the ray-tracer output. The ray-tracer reports the paths between transmitters and receivers and calculates \gls{toa}, received power, and phase shift of the received signal, as well as the angular characteristics for each path. 
This process takes into account the path trajectory distance and the reflection coefficient of the materials at each reflection point. In \gls{wi}, the simulation output finds the paths with power $> -250\:\mathrm{dBm}$, which include the ones below the noise floor.
Then, we prune the paths with received power lower than the noise floor. This is computed using Equation~\ref{eq:noise_level}, where $N_o$, $B$ and $F$ are the ambient noise density $[dBm/Hz]$, the receiver bandwidth $[Hz]$, and the receiver noise figure $[dB]$, respectively.
\begin{equation}
    \label{eq:noise_level}
    Noise [dBm] = N_o + 10 * \log B + F
\end{equation}
 However, \gls{wi} does not directly report the gain parameter of the paths for the valid channels at time instant $t$, so we calculate the gain of the paths as complex numbers using Equation~\ref{eq:paths_gain}, where $P_{Tx}$ is transmit power in $dB$, and $P_{Rx_i}$ and $\varphi_i$ are the received power and signal phase per each path $i\in\{1..N_t\}$, respectively. 
\begin{equation}
    \label{eq:paths_gain}
    \Tilde{c_i}(t) = 10^{(P_{Rx_i}(t) - P_{Tx})/20} * e^{j\varphi_i(t)} \qquad i = 1 .. N_t
\end{equation}

We simulate the use-case scenario discussed in Section~\ref{sec:usecase_scenario} in \gls{wi} by considering 4 reflections to find the paths between the transmitter and receivers for a reasonable simulation time.
We compute the time-variant \gls{cir} of the \gls{rsu} to \gls{obu}\#3 channel, which is depicted later in Figure~\ref{fig:mobileorigpaths} (see Section~\ref{sec:results}).
As a final metric, we can evaluate the impact of mobility on path loss, which is one of the important channel parameter for large-scale channel characterization to recognize the fading impact of mobility and multi-paths~\cite{molisch2005ultrawideband}. To this end, we calculate $L_p$, link path loss, using Equation~\ref{eq:link_pathloss} which is the magnitude of coherently summation of the coefficients in dB. The results of path loss are analyzed in details in Section~\ref{sec:resultsmobility}.
\begin{equation}
    \label{eq:link_pathloss}
    L_p(t) = -20.\log \left | \sum_{i=1}^{N_t} \Tilde{c_i}(t) \right |
\end{equation}

\subsection{Emulation channel taps approximation}
Due to computational restrictions and trade-offs, \gls{fir}-based channel emulators can only account for limited number of non-zero filter taps (4 taps in the case of Colosseum \gls{mchem}~\cite{ChaudhariST18}). However, the ray-tracer derived models typically include numerous paths between transmitters and receivers. Furthermore, delays of paths may not be necessarily aligned with the pre-defined step-wise delays of the emulator \gls{fir} filter taps.
%
To this end, we use the \emph{tap approximation} framework proposed in our recent work~\cite{tehrani2021creating} that employs a \gls{ml}-based clustering method to convert the ray-tracer channels into the taps format compatible with the requirements/limitations of the channel emulator. This involves approximating the channel to an acceptable number of taps, aligning the tap delays to the pre-defined indices, and adjusting the dynamic range of the taps while preserving a desirable accuracy from the original channel.

\section{Experimental Results}
\label{sec:results}
We first run \gls{cast} in a lab testbed (Figure~\ref{fig:localtestbed}) to tune the parameters in a controlled environment in the absence of channel emulator impairments. One of the main goals of this step is to find a code sequence that can result in a high auto-correlation and a low cross-correlation between transmitted code sequence and received signal, which consequently can reveal channel taps. Secondly, \gls{cast} is used to understand the behavior of Colosseum emulation by testing a set of synthetic scenarios, i.e., created specifically for the sounding purpose. Finally, quasi-real-world scenarios with static and mobile nodes developed in Section~\ref{sec:mobility} are deployed and characterized.

A customized \gls{lxc} image containing \gls{cast} is created and uploaded to Colosseum. This container has all the required libraries and software for the channel sounding system and its post-processing operations. This enables the re-usability of the sounding with different Colosseum \gls{srn} and scenarios, and allows for the automation of all the processes till the generation of the final results. This image is open-source and will be made available for all Colosseum users in the publicly accessible \gls{nas}.

\subsection{Lab testbed validation}

\begin{figure}[hbp]
\centering
    \centering
    \includegraphics[width=0.9\linewidth]{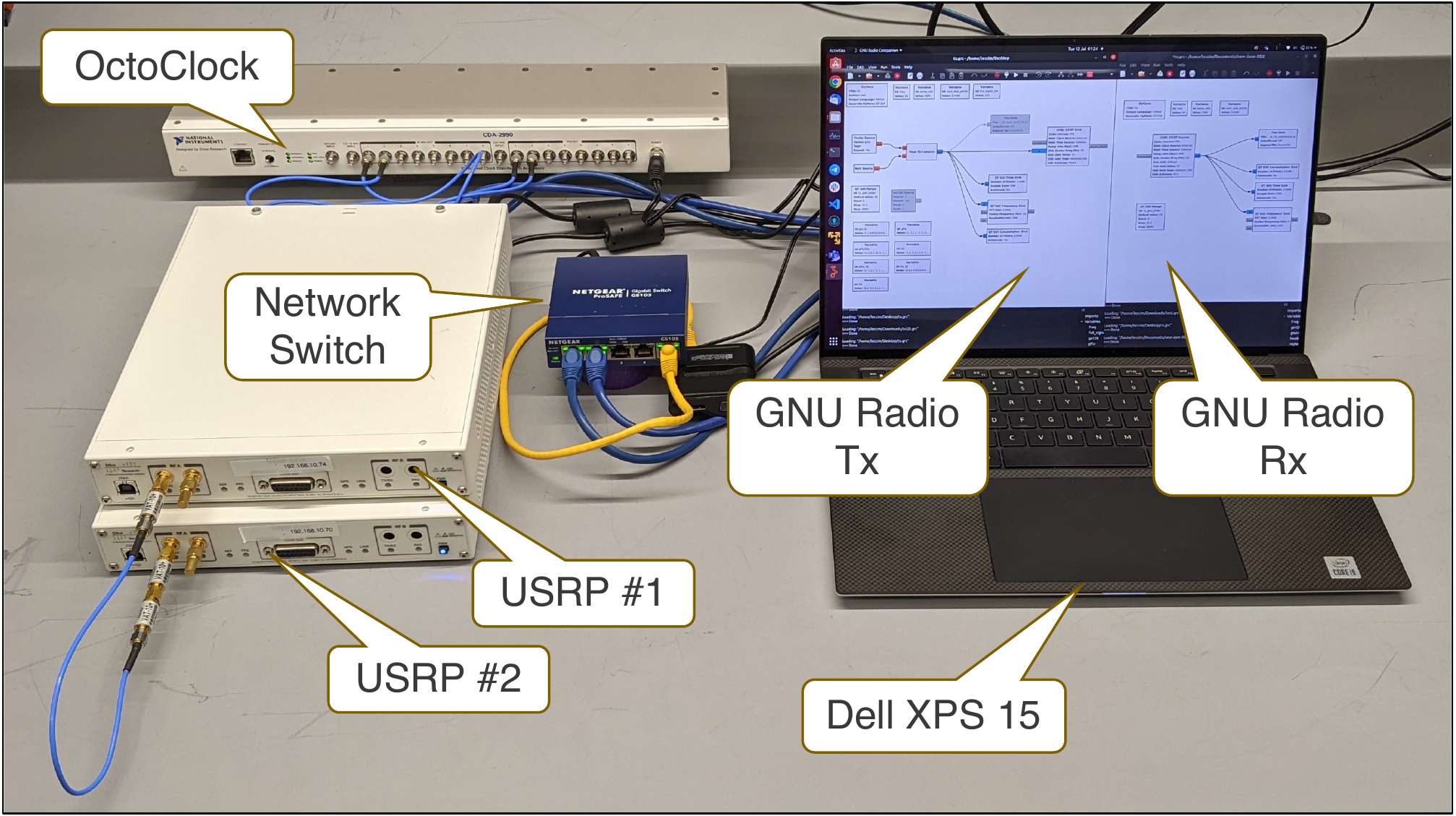}
    \vspace{-5pt}
    \caption{Lab testbed environment.}
    \label{fig:localtestbed}
\end{figure}

The lab testbed is shown in Figure~\ref{fig:localtestbed}.

It consists of two NI/Ettus Research \gls{usrp} X310 radios, each one equipped with one UBX-160 daughterboard. 
The radios are synchronized in time and frequency through an NI/Ettus OctoClock clock distributor CDA-2990 which generates the clock internally. A Dell XPS 15 computer is used to program the radios, and to perform the post-processing operations.
The default connection between the two \gls{usrp} consists of a $12\:\mathrm{inches}$ SMA cable and three attenuators for a total of $30\:\mathrm{dB}$ of attenuation. 
The sounding parameters for the lab experiments are summarized in Table~\ref{table:localconfig}. The gains of the \gls{usrp} vary between $0$ and $15\:\mathrm{dB}$ to understand their effect. The code sequence is a \gls{glfsr} (the effectiveness of different code sequences are analysed in Section~\ref{sec:codeseq}). The receiving period time and data acquisition are set to $3\:\mathrm{s}$.

\begin{figure}[htbp]
\centering
    \centering
    \includegraphics[width=.99\linewidth]{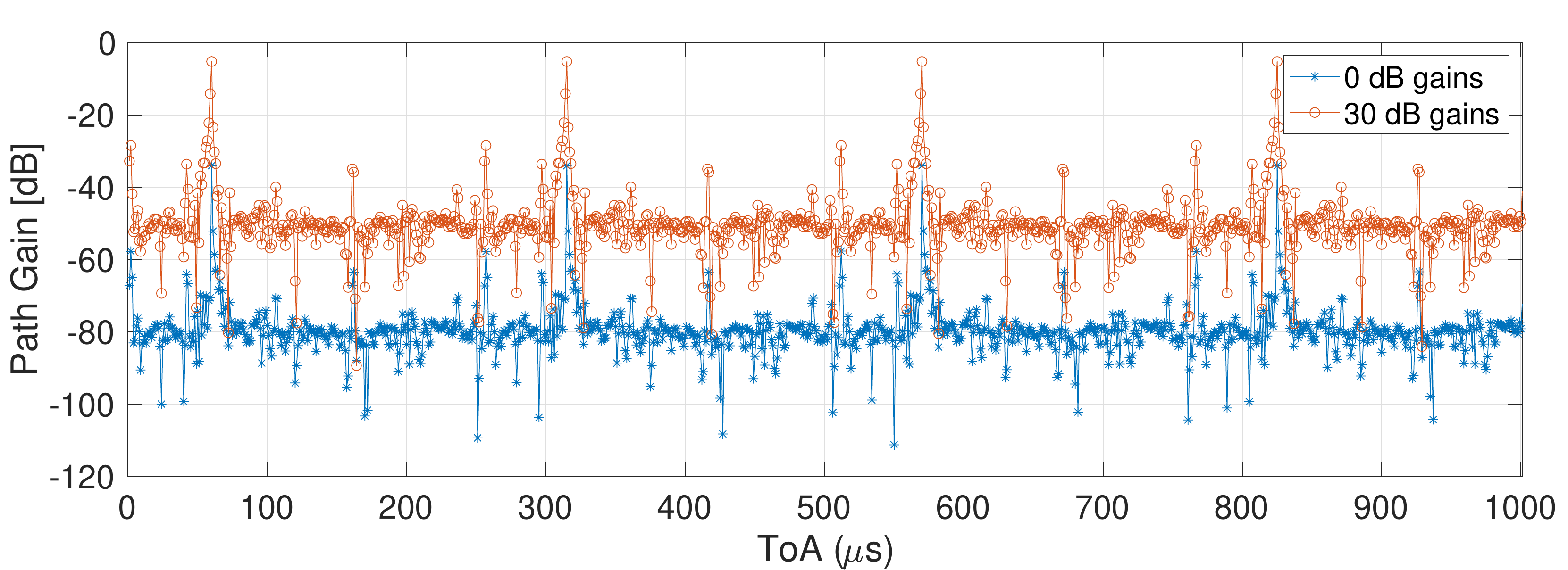}
    \vspace{-5pt}
    \caption{Received path gains lab testbed in $0$ and $30\:\mathrm{dB}$ gains use cases.}
    \label{fig:pglocal}
\end{figure}

Figure~\ref{fig:pglocal} shows a time frame of the received path gains for the case with $0\:\mathrm{dB}$ gains (blue line) and $30\:\mathrm{dB}$ gains, consisting of $15\:\mathrm{dB}$ on the transmitter and $15\:\mathrm{dB}$ on the receiver (orange line). 

The signal cycles based on the transmitted sequence length, i.e., every 255 sample points, or equivalently every $255\:\mathrm{\mu s}$ since 1 point is equal to $1/sample\_rate = 1\:\mathrm{\mu s}$. The peaks represent the path loss of the single tap of this lab experiment and is equal to $34.06\:\mathrm{dB}$ for the $0\:\mathrm{dB}$ case and $5.24\:\mathrm{dB}$ for the $15\:\mathrm{dB}$. These numbers are in line with the expectation since we have $30\:\mathrm{dB}$ loss due to the attenuators and a few more due to the radios, cable, attenuators, computational imprecision, and some background noise. Moreover, we can notice that in the $30\:\mathrm{dB}$ case, the loss is slightly more since there \gls{usrp} might not add $30\:\mathrm{dB}$ total but slightly less. These numbers are used as a reference point for the next experiments and validations.

\begin{table}[hbp]
\centering\caption{Configuration parameters for the lab testbed.}
\begin{tabular}{@{}ll@{}}
\toprule
\multicolumn{1}{c}{\textbf{Parameter}} & \multicolumn{1}{c}{\textbf{Value}} \\ \midrule
Center frequency & $1\:\mathrm{GHz}$ \\
Sample rate & Various $1-50\:\mathrm{MS/s}$ \\
\gls{usrp} tx gain & Various $0-15\:\mathrm{dB}$ \\
\gls{usrp} rx gain & Various $0-15\:\mathrm{dB}$ \\
Code sequence & \gls{glfsr} $255\:\mathrm{bits}$ \\ \bottomrule
\end{tabular}
\label{table:localconfig}
\end{table}


\subsection{Code sequences}\label{sec:codeseq}
Code sequences have been widely investigated in literature given their high utility in many different fields~\cite{velazquez2016sequence}~\cite{stanczak2001are}. Good code sequences target a high auto-correlation, i.e., correlation between two copy of the same sequence, and low cross-correlation, i.e., correlation between two different sequences.
By exploiting the lab testbed environment described above, four code sequences are tested to find out the one with the \gls{cir} that best fits our experiments:
\begin{enumerate}[$\bullet$]
    \item Gold Sequence~\cite{zhang2011analysis} of $255\:\mathrm{bits}$ created with the Matlab Gold sequence generator System object with its default parameters.
    \item Golay Sequence type A (Ga$_{128}$)~\cite{ieee-802_11ad-standard} with a size of $128\:\mathrm{bits}$ and defined in the IEEE Standard.
    \item \gls{ls} Code~\cite{perez2007efficient} generated via~\cite{garcia2010generation} exploiting just the first codeset of $\{1,-1\}$ without including the \gls{ifw}.
    \item \gls{glfsr}~\cite{pradhan1999glfsr} of $255\:\mathrm{bits}$ created via the GNU Radio \textit{GLFSR Source} pseudo-random generation block with the following parameters: degree of shift register 8, bit mask 0, and seed 1.
\end{enumerate}
\begin{figure}[bp]
    \centering
    \begin{subfigure}[b]{0.475\columnwidth}
        \includegraphics[width=\columnwidth]{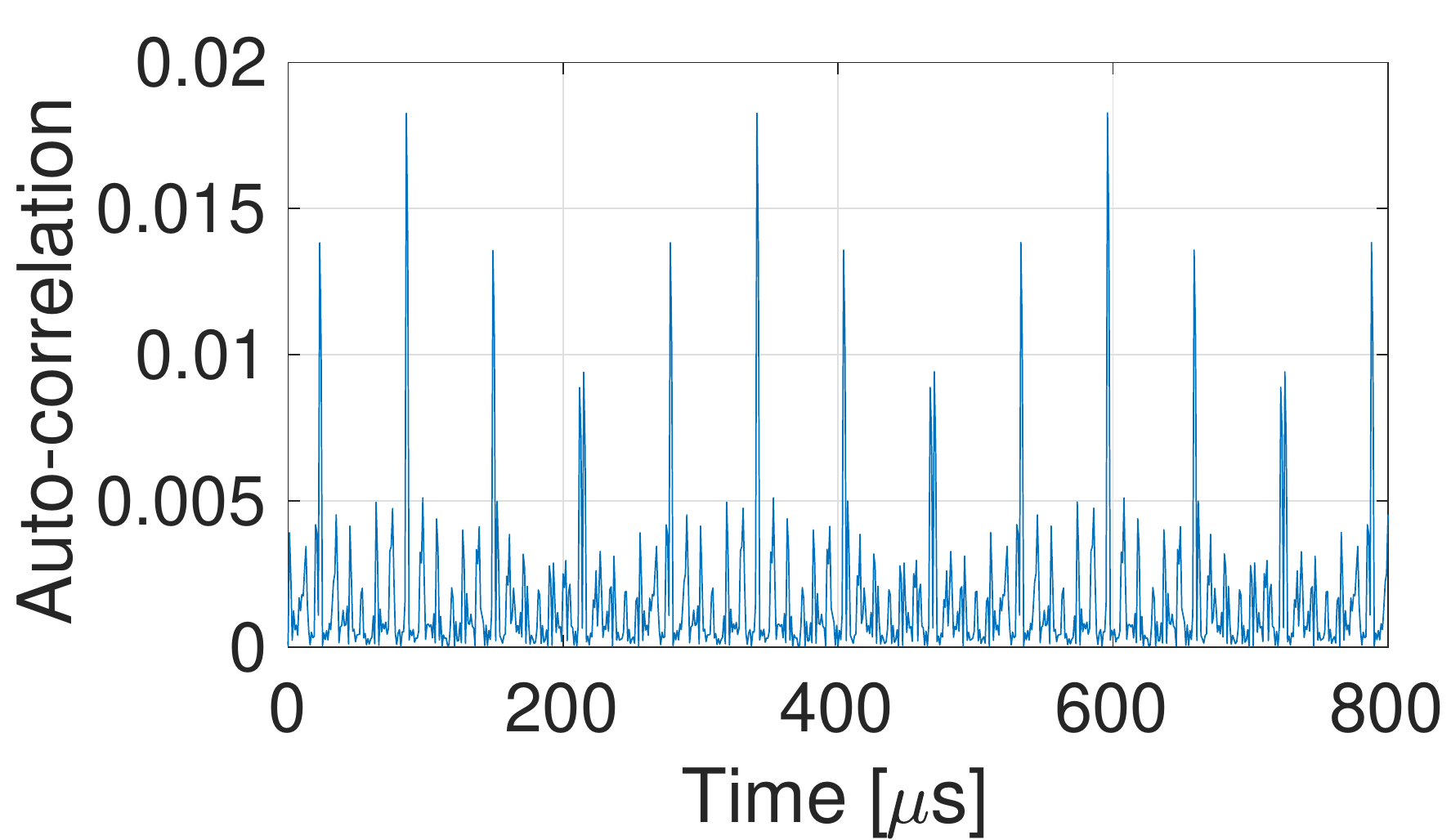}
        \caption{Gold Sequence.}
        \label{fig:cirlocalgold}
    \end{subfigure}
    \begin{subfigure}[b]{0.475\columnwidth}
        \includegraphics[width=\columnwidth]{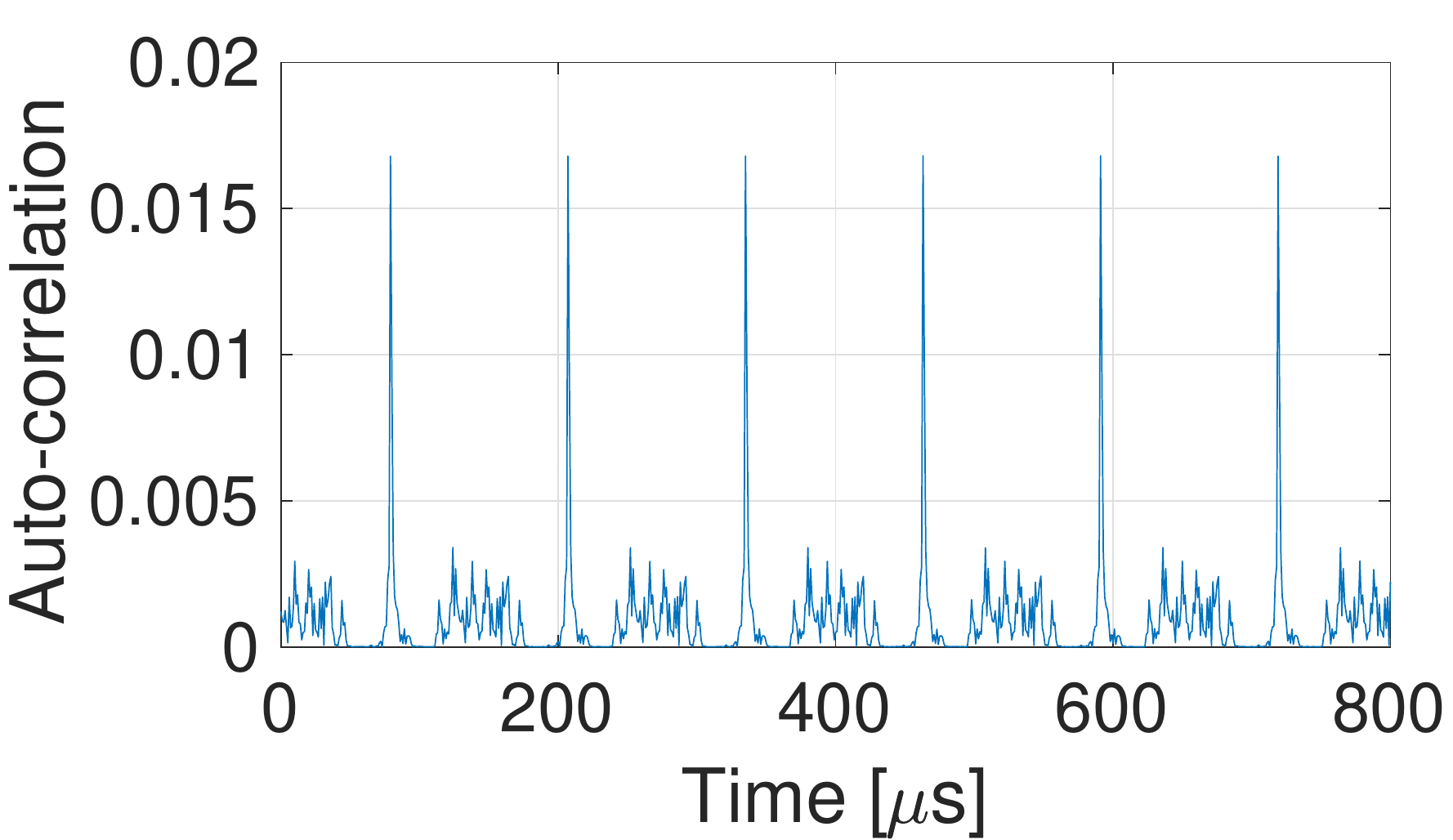}
        \caption{Ga$_{128}$ Code.}
        \label{fig:cirlocalga}
    \end{subfigure}
    \begin{subfigure}[b]{0.475\columnwidth}
        \includegraphics[width=\columnwidth]{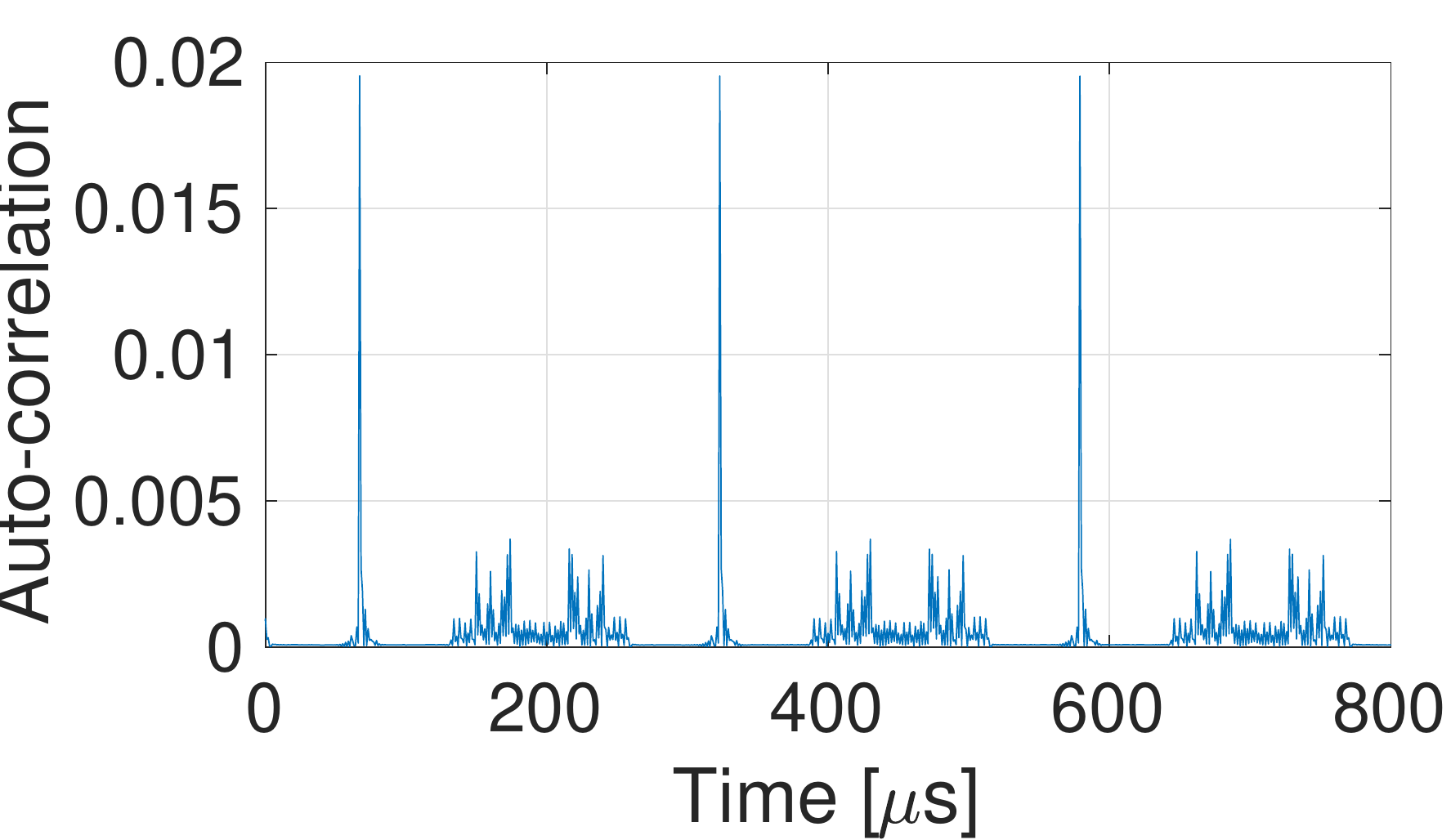}
        \caption{\gls{ls} Code.}
        \label{fig:cirlocalls1}
    \end{subfigure}
    \begin{subfigure}[b]{0.475\columnwidth}
        \includegraphics[width=\columnwidth]{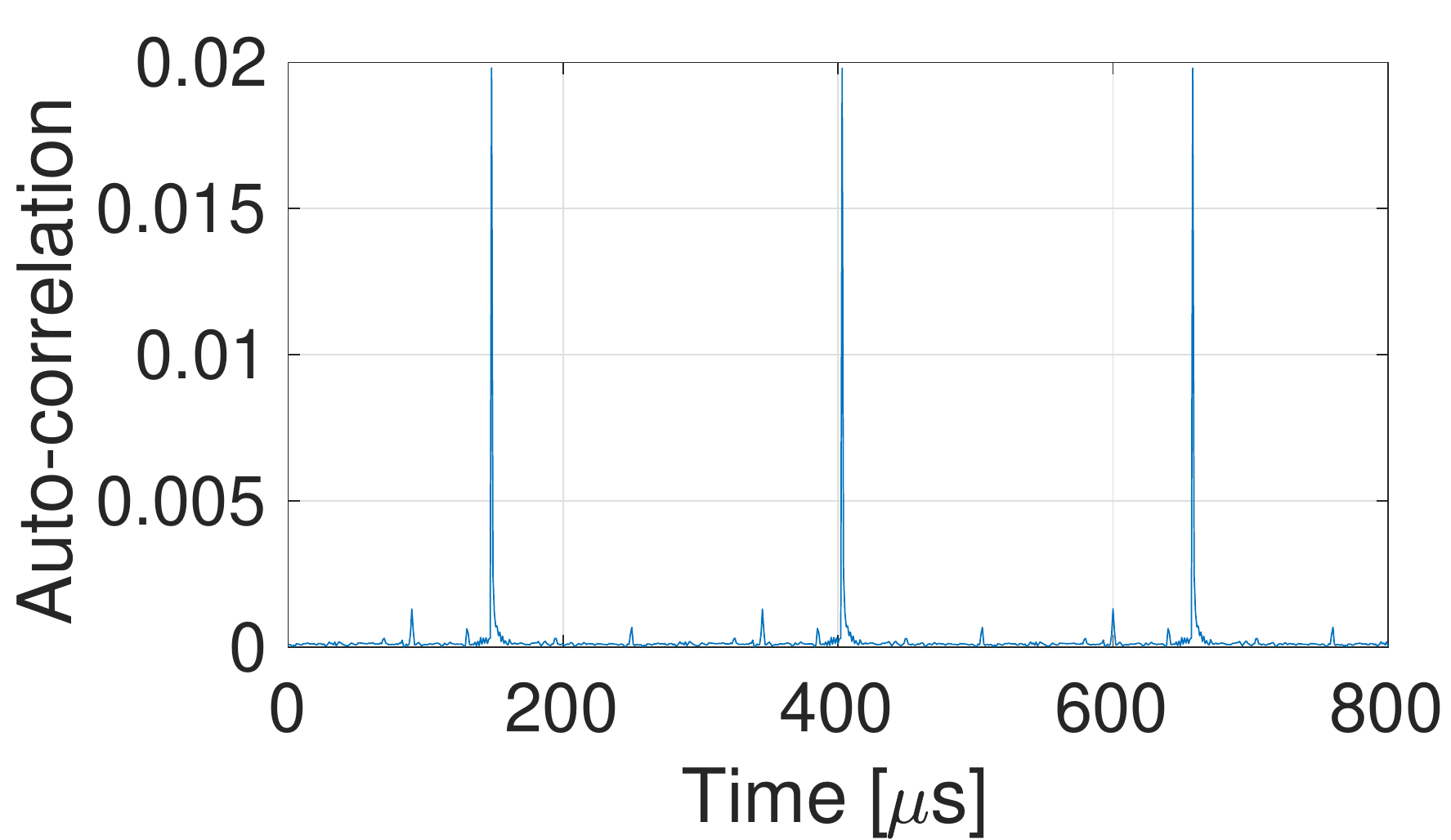}
        \caption{\gls{glfsr} Code.}
        \label{fig:cirlocalglfsr}
    \end{subfigure}
    \caption{\gls{cir} for different code sequences in the lab testbed.}
    \label{fig:cirlocal}
\end{figure}
The results of a time frame \gls{cir} for each code sequence is show in Figure~\ref{fig:cirlocal}. We can see that the \gls{glfsr} sequence outperforms the other three having the highest auto-correlation, and lower cross-correlations compared to the other ones resulting in a cleaner \gls{cir}. For this reason, in our experiments the \gls{glfsr} is used as the main code sequence source.

\subsection{Synthetic scenario validation}
The first set of scenarios are synthetic, i.e., manually generated with specific characteristics to validate \gls{mchem} behavior. The set of used parameters is the same as Table~\ref{table:localconfig} but with a sample rate of $50\:\mathrm{MS/s}$ to have enough resolution ($20\:\mathrm{ns}$ per sample) to properly retrieve tap delays and gains.

\begin{figure}[htbp]
\centering
    \centering
    \includegraphics[width=0.99\linewidth]{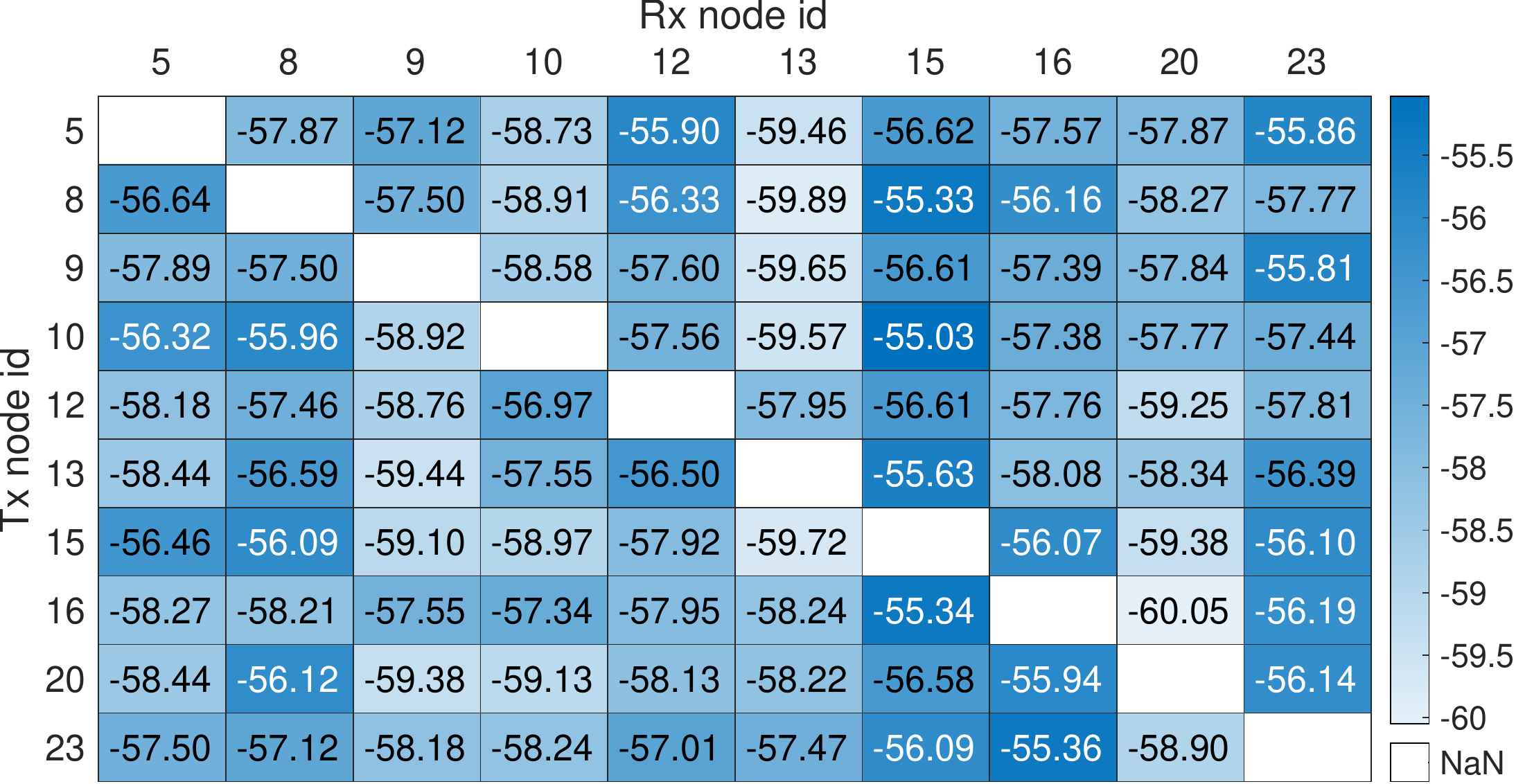}
    \caption{Path loss heatmap in a $0\:\mathrm{dB}$ scenario with~10 nodes.}
    \label{fig:heatmap0db}
\end{figure}

The first tested scenario is the simplest one single-tap with nominal $0\:\mathrm{dB}$ path loss. The results show that the signal properly cycles every $5.1\:\mathrm{\mu s}$ with a recurrent average loss of around $58\:\mathrm{dB}$. This loss can be traced back to Colosseum equipment in-the-loop, consisting of four \gls{usrp} X310 radios and several cables, and emulation computation approximations. However, the origin of this loss will require further investigation. We can refer to this number as Colosseum base loss.
To confirm its value, Figure~\ref{fig:heatmap0db} shows the results for 10 nodes in the same $0\:\mathrm{dB}$ scenario. 

Each cell represents the average path loss for $2\:\mathrm{s}$ reception time between a transmitter node (row id) and a receiver node (column id) using both the first daugherboard. 
Results confirm an average Colosseum base loss of $57.55\:\mathrm{dB}$ with a \gls{sd} of $1.23\:\mathrm{dB}$.
We also observe that the current dynamic range of Colosseum operations is around $43\:\mathrm{dB}$, i.e., between the maximum value at $0\:\mathrm{dB}$ scenario of $57.55\:\mathrm{dB}$ and the minimum one, given by the noise floor at~$100\:\mathrm{dB}$.
These are fundamental findings that need to be considered when designing scenarios and analysing results.

\begin{figure}[htbp]
\centering
    \centering
    \includegraphics[width=.99\linewidth]{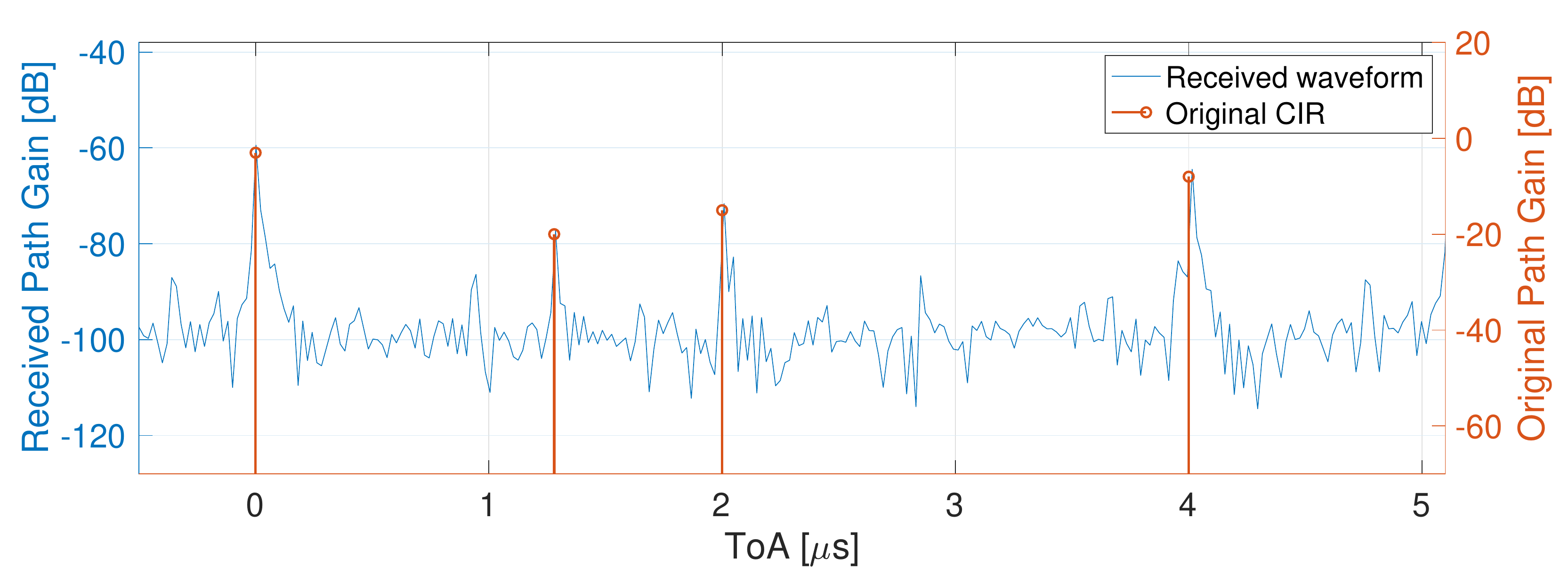}
    \caption{Received path gains and original \gls{cir} comparison.}
    \label{fig:cirpgcomparison}
\end{figure}

The next synthetic scenario analyzed consists of four taps with different delay times and path gains (Figure~\ref{fig:cirpgcomparison} orange stems). The resulting received gains (Figure~\ref{fig:cirpgcomparison} blue lines) show that the \gls{toa} of the different taps are exactly the same, namely $[0 - 1.28 - 2 - 4]\:\mathrm{\mu s}$.

The received powers are in line with the expectations. If we add the Colosseum base loss, the gain results (right y-axis) match the original ones (left x-axis), particularly $[3 - 20 - 15 - 8]\:\mathrm{dB}$ losses.
By considering a large number of time frames, e.g., fifteen hundreds, we can calculate the relative differences of the received taps over time to obtain information on the accuracy of these measurements. The average difference between the highest first taps of each time frame is in the order of $10^{-6}$ with a \gls{sd} of $0.03\:\mathrm{dB}$. The same average happens for the lowest second taps with a slightly higher \gls{sd} of~$0.17\:\mathrm{dB}$. On the other hand, considering the differences between the highest (1st tap) and lowest (2nd tap) for each time frame, we have an average difference of $0.52\:\mathrm{dB}$ with a \gls{sd} of $0.18\:\mathrm{dB}$. These results are a consequence of the contribution of the noise, which is largest in the lowest taps compared to the highest ones.
These results prove that \gls{mchem} is emulating the channel correctly for both delay and gain taps, and that the received expected signal is consistent with the original one.
Moreover, this shows that \gls{cast} is able to achieve a resolution of $20\:\mathrm{ns}$, sustaining the sample rate of $50\:\mathrm{MS/s}$ and an accuracy on the tap gains measurements of $0.5\:\mathrm{dB}$.

\begin{figure}[bp]
    \centering
    \begin{subfigure}[b]{\columnwidth}
        \includegraphics[width=0.95\columnwidth]{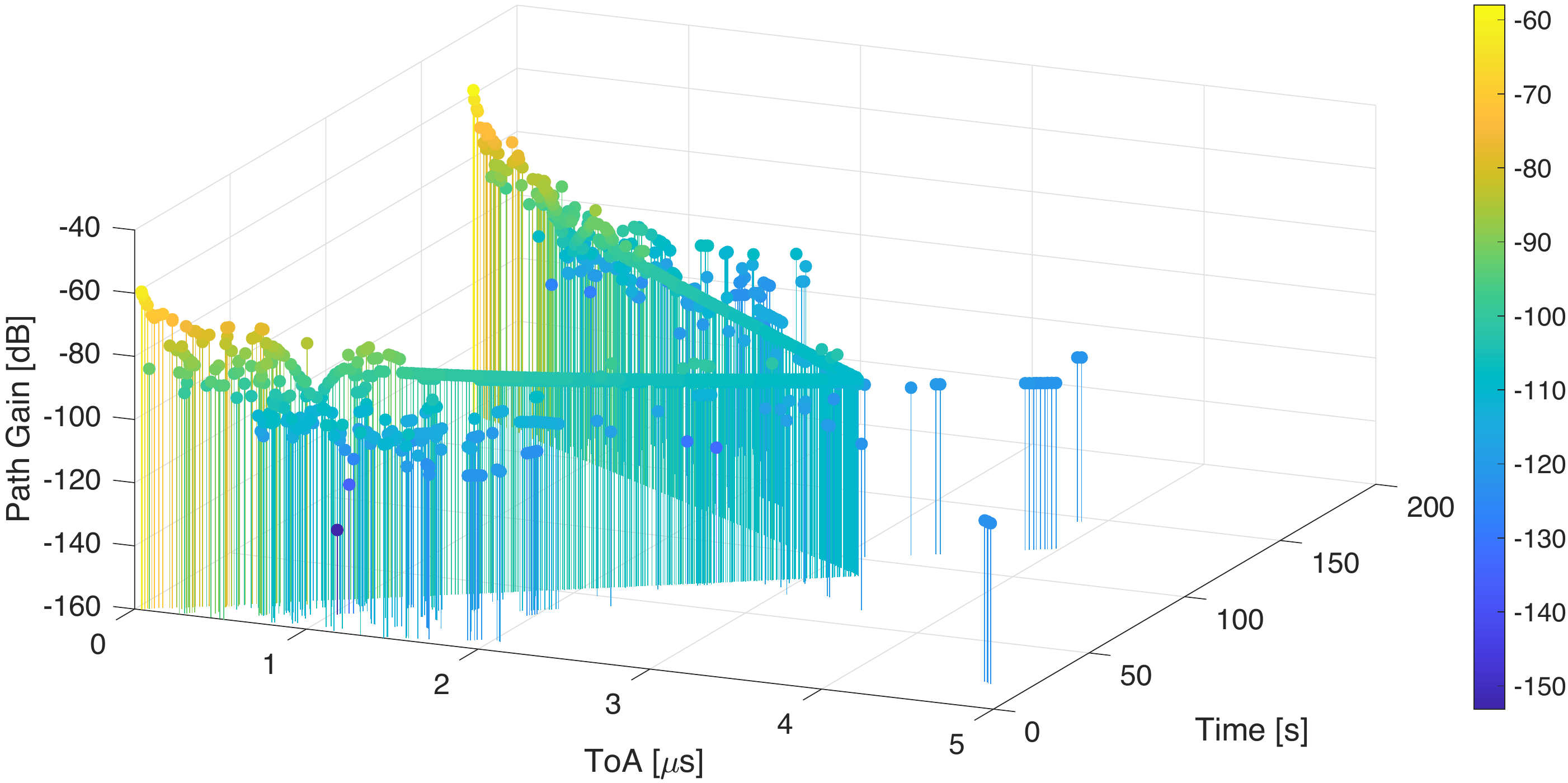}
        \caption{Original taps representation.}
        \label{fig:mobileorigpaths}
    \end{subfigure} \\
    \begin{subfigure}[b]{\columnwidth}
        \includegraphics[width=0.95\linewidth]{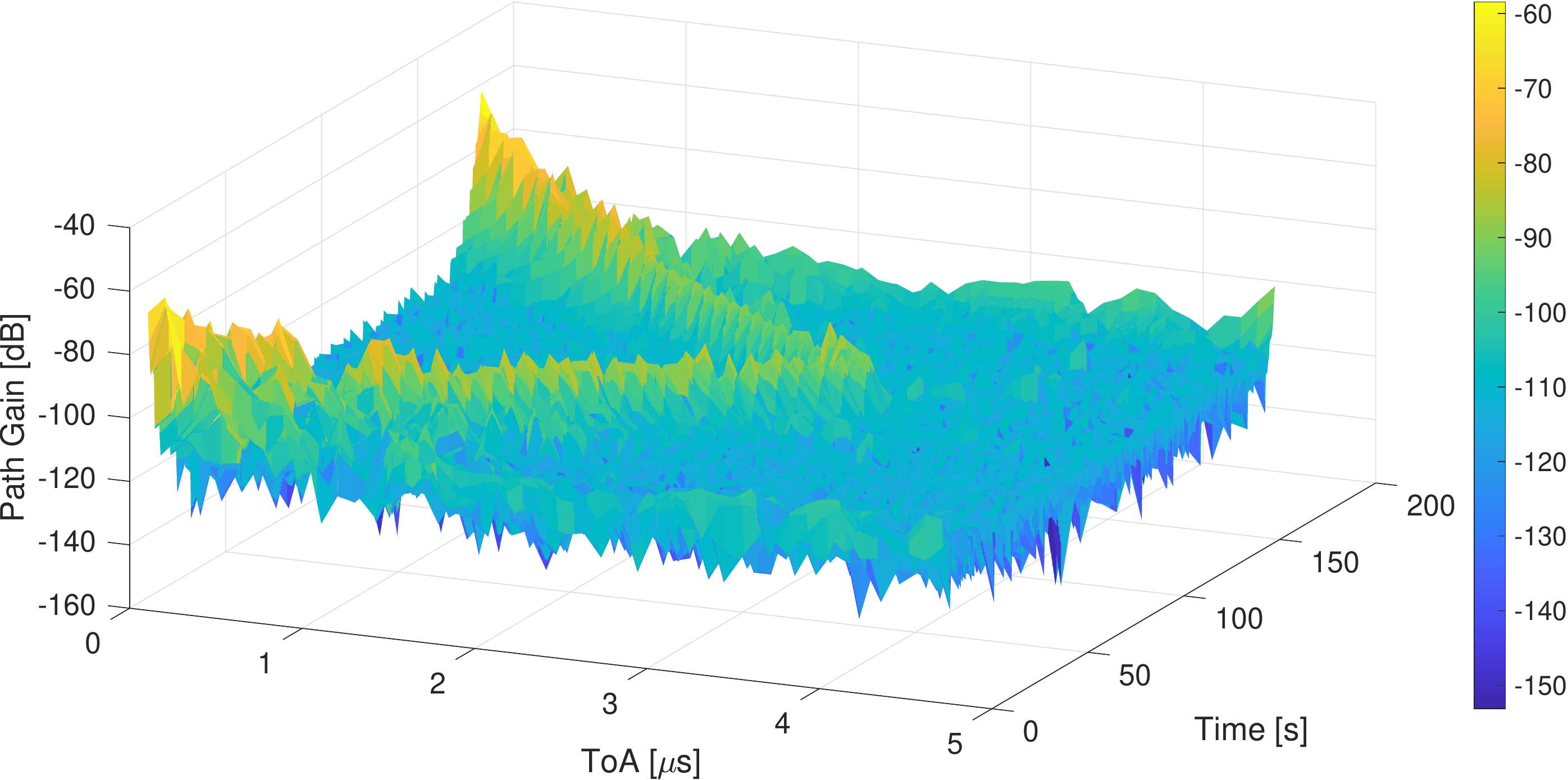}
        \caption{Received path gains.}
        \label{fig:mobileallpg}
    \end{subfigure}
    \caption{Results comparison for the mobile use case scenario.}
    \label{fig:3dorigmobile}
\end{figure}

\subsection{Mobility scenario validation}
\label{sec:resultsmobility}

\begin{figure}[htbp]
\centering
    \centering
    \includegraphics[width=0.99\linewidth]{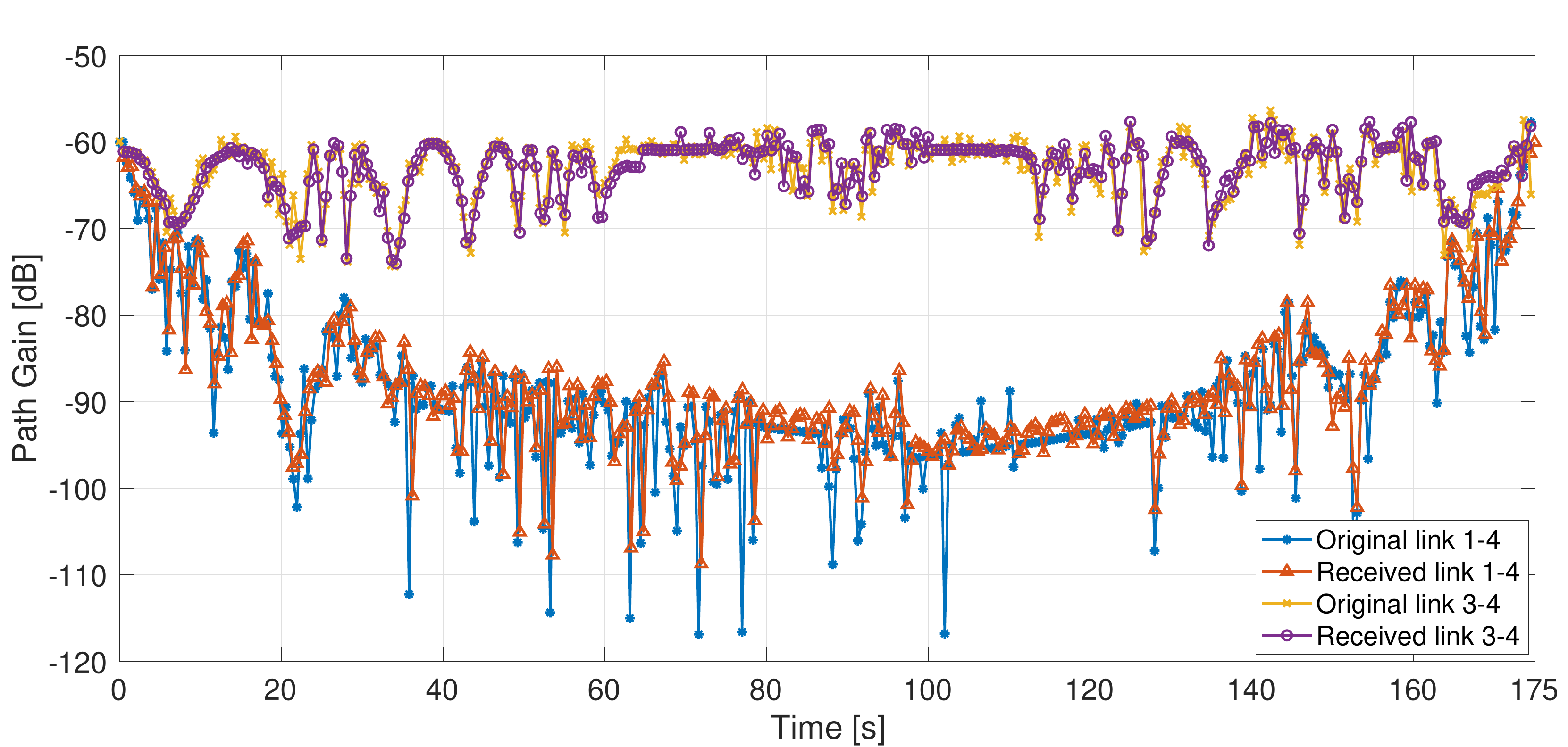}
    \caption{Comparison between original and received path gains for \gls{obu}\#3 (node 4) against \gls{rsu} (node 1) and \gls{obu}\#2 (node 3).}
    \label{fig:recorigpathloss}
\end{figure}

In the last set of experiments we test the mobility scenario described in Section~\ref{sec:usecase_scenario}. 
This scenario has been installed in Colosseum with an increase of $60\:\mathrm{dB}$ in all taps, to fall inside the Colosseum dynamic range.
The parameters for the sounding process are the same as those listed in Table~\ref{table:localconfig} with $15\:\mathrm{dB}$ gains and $10\:\mathrm{MS/s}$ sample rate. 
Since the total scenario time is $175\:\mathrm{s}$ and processing all the data together would require extreme memory, the rx time is divided into three chunks of around $60\:\mathrm{s}$ each. In this way, each chunk is around $5\:\mathrm{GB}$ in size and it takes about $30\:\mathrm{minutes}$ to be processed. The results of each chunk are cleaned and merged together to create the ultimate outcome. The received path gains have been adjusted by removing the Colosseum base loss and adding the original $60\:\mathrm{dB}$ increase.
Figure~\ref{fig:mobileallpg} shows how the received path gains (xy-axis) vary over the scenario time (z-axis). We can notice that the strongest tap resembles the same `V-shape' behavior seen in the original taps representation (Figure~\ref{fig:mobileorigpaths}) as a direct consequence of the movement to and from the static \gls{rsu} node.
Moreover, Figure~\ref{fig:recorigpathloss} shows the link path loss of node 1 (\gls{rsu}) and mobile node 3 (\gls{obu}\#2) against the mobile node 4 (\gls{obu}\#3). 

The overall original and received results are fluctuating due to the multi-path fading, but they almost perfectly align between each other. In this case, for link 1-4 a very similar received power trend with a convex `U-shape' is noticeable. The mobile node is first getting away from the \gls{rsu} moving south decreasing the gains and increasing the \gls{toa}, then  the pattern is reserved as the vehicle turns around and travels back to the intersection and the \gls{rsu}. On the other hand, link 3-4 has a more stable trend since the two vehicle nodes perform the journey together.
These results confirm that Colosseum is emulating the channel correctly even in a mobility scenario, and validates the capabilities of \gls{cast} even when the metrics are changing over time.

\section{Conclusions}
\label{sec:conclusions}

This paper presents \gls{cast}, a fully open, publicly available, software-based and virtualized realistic channel generator and sounder tool\-chain. 
It enables researchers to design, create, and validate realistic channel models in scenarios with mobile nodes.
The system is validated through experimental tests on a lab setup and on Colosseum by designing, deploying and validating multi-path mobile scenarios. 
Our results show that \gls{cast} achieves up to $20\:\mathrm{ns}$ accuracy in sounding the \gls{cir} tap delays, and $0.5\:\mathrm{dB}$ accuracy in measuring the tap gains.
Future work on \gls{cast} include enhancing current results by adding signal propagation delay and phase sounding; accelerating data processing operations by using new FPGA and GPU-based solutions; and improving accuracy by considering techniques that reduce noise and fluctuations.



\bibliographystyle{ACM-Reference-Format}
\bibliography{biblio}

\end{document}